\documentclass[twocolumn]{aastex61}

\begin{document}

\shorttitle{The Isophotal Structure of SFGs at $0.5<\lowercase{z}<1.8$ in CANDELS}
\title{The Isophotal Structure of Star-forming Galaxies at $0.5<\lowercase{z}<1.8$ 
in CANDELS: Implications for the Evolution of Galaxy Structure}

\correspondingauthor{F. S. Liu}
\email{Email: fengshan.liu@yahoo.com}

\author{Dongfei Jiang} 
\affil{College of Physical Science and Technology, Shenyang Normal University, Shenyang 110034, China}
\affiliation{Purple Mountain Observatory, Chinese Academy of Sciences,
2 West-Beijing Road, Nanjing 210008, China}

\author{F. S. Liu $^{\color{blue} \dagger}$}
\affil{College of Physical Science and Technology, Shenyang Normal University, Shenyang 110034, China}
\affiliation{University of California Observatories and the Department of Astronomy and Astrophysics, 
University of California, Santa Cruz, CA 95064, USA}

\author{Xianzhong Zheng}
\affiliation{Purple Mountain Observatory, Chinese Academy of Sciences,
2 West-Beijing Road, Nanjing 210008, China}
\affiliation{Chinese Academy of Sciences South America Center for Astronomy,
China-Chile Joint Center for Astronomy, Camino El Observatorio 1515, Las Condes, Santiago, Chile}

\author{Hassen M. Yesuf}
\affiliation{University of California Observatories and the Department of Astronomy and Astrophysics,
University of California, Santa Cruz, CA 95064, USA}

\author{David C. Koo}
\affiliation{University of California Observatories and the Department of Astronomy and Astrophysics,
University of California, Santa Cruz, CA 95064, USA}

\author{S. M. Faber}
\affiliation{University of California Observatories and the Department of Astronomy and Astrophysics,
University of California, Santa Cruz, CA 95064, USA}

\author{Yicheng Guo}
\affiliation{Department of Physics and Astronomy, University of Missouri, Columbia, MO 65211, USA}
\affiliation{University of California Observatories and the Department of Astronomy and Astrophysics,
University of California, Santa Cruz, CA 95064, USA}

\author{Anton M. Koekemoer}
\affil{Space Telescope Science Institute, 3700 San Martin Drive, Baltimore, MD 21218, USA}

\author{Weichen Wang}
\affiliation{Department of Physics \& Astronomy, Johns Hopkins University, 3400 N. Charles Street, Baltimore, MD 21218, USA}

\author{Jerome J. Fang}
\affiliation{University of California Observatories and the Department of Astronomy and Astrophysics,
University of California, Santa Cruz, CA 95064, USA}
\affiliation{Astronomy Department, Orange Coast College, Costa Mesa, CA 92626, USA}

\author{Guillermo Barro}
\affiliation{University of California Observatories and the Department of Astronomy and Astrophysics,
University of California, Santa Cruz, CA 95064, USA}
\affiliation{Department of Astronomy, University of California, Berkeley, CA 94720-3411, USA}
`
\author{Meng Jia}
\affil{College of Physical Science and Technology, Shenyang Normal University, Shenyang 110034, China}

\author{Wei Tong}
\affil{College of Physical Science and Technology, Shenyang Normal University, Shenyang 110034, China}

\author{Lu Liu}
\affil{College of Physical Science and Technology, Shenyang Normal University, Shenyang 110034, China}

\author{Xianmin Meng}
\affil{National Astronomical Observatories, Chinese Academy of Sciences, A20 Datun Road, Beijing 100012, China}

\author{Dale Kocevski}
\affil{Department of Physics and Astronomy, Colby College, Mayflower Hill Drive, Waterville, ME 0490, USA}

\author{Elizabeth J. McGrath}
\affil{Department of Physics and Astronomy, Colby College, Mayflower Hill Drive, Waterville, ME 0490, USA}

\author{Nimish P. Hathi}
\affil{Aix Marseille Universit\'e, CNRS, LAM (Laboratoire d'Astrophysique de Marseille) UMR 7326, 13388, Marseille, France}
\affil{Space Telescope Science Institute, 3700 San Martin Drive, Baltimore, MD 21218, USA}

\begin{abstract}

We have measured the radial profiles of isophotal
ellipticity ($\varepsilon$) and disky/boxy parameter A$_4$
out to radii of about three times the semi-major axes for $\sim4,600$
star-forming galaxies (SFGs) at intermediate redshifts $0.5<\lowercase{z}<1.8$ in the CANDELS/GOODS-S and UDS fields.
Based on the average size versus stellar-mass relation in each redshift bin,
we divide our galaxies into \emph{Small} SFGs (SSFGs), i.e., smaller than
average for its mass, and \emph{Large} SFGs (LSFGs), i.e., larger than average.
We find that, at low masses ($M_{\ast} < 10^{10}M_{\odot}$), the SSFGs generally have nearly flat $\varepsilon$
and A$_4$ profiles for both edge-on and face-on views, especially at redshifts $z>1$.
Moreover, the median A$_4$ values at all radii are almost zero.
In contrast, the highly-inclined, low-mass LSFGs in the same mass-redshift bins
generally have monotonically increasing $\varepsilon$ with radius
and are dominated by disky values at intermediate radii.
These findings at intermediate redshifts imply that low-mass SSFGs are not disk-like, while
low-mass LSFGs appear to harbour disk-like components flattened by significant rotation.
At high masses ($M_{\ast} > 10^{10}M_{\odot}$), highly-inclined SSFGs and LSFGs
both exhibit a general, distinct trend for both $\varepsilon$ and A$_4$ profiles:
increasing values with radius at lower radii, reaching maxima at intermediate radii, and then  decreasing values at larger radii.
Such a trend is more prevalent for more massive ($M_{\ast} > 10^{10.5}M_{\odot}$)
galaxies or those at lower redshifts ($z<1.4$).
The distinct trend in $\varepsilon$ and A$_4$ can be simply explained if galaxies possess all three components:  central bulges,
disks in the intermediate regions, and halo-like stellar components in the outskirts.

\end{abstract}

\keywords{galaxies: photometry --- galaxies: star formation --- galaxies:
high-redshift}

\section{Introduction}

In the $\Lambda$CDM framework of hierarchical growth of structures, 
galaxies are assembled by mergers and low-mass accretion events 
\cite[e.g.][]{Eggen1962ApJ,Sandage1986ARA,Purcell2007ApJ,Johnston2008ApJ,
DeLucia2008MNRAS,Cooper2013MNRAS,Pillepich2014MNRAS}. 
While disk galaxies can be formed in the centers of 
dark-matter halos via gas infall, elliptical galaxies and 
bulges, which we refer to as spheroids, form via violent major mergers 
\cite[e.g.][]{Kauffmann1993MNRAS,Baugh1996MNRAS}.
Stars stripped from infalling satellite galaxies 
form a diffuse and highly-structured stellar halo 
surrounding the central galaxy.  As a consequence of the relatively-long, dynamical timescales in the outskirts of galaxies, 
such halos retain a ``memory'' of past accretion events
\cite[e.g.][]{Eggen1962ApJ,Searle1978ApJ...225..357S,Steinmetz1995MNRAS.276..549S,
Bekki2001ApJ...558..666B,Samland2003A&A...399..961S}.

The past decade has experienced major advances in our understanding of 
the formation and evolution of bulges \citep[see][for reviews]{Tonini2016MNRAS.459.4109T,Tonini2016arXiv160606040T}.
Bulges are now divided into two main types: classical bulges and pseudo-bulges 
\cite[][]{Kormendy1993IAUS,Kormendy04}.
Classical bulges resemble elliptical galaxies by being dynamically-hot spheroids 
with stellar motions dominated by velocity dispersion rather than rotation; they usually have 
a strongly-concentrated structure (e.g., S\'ersic index n$\sim$4).
The pseudo-bulges are dynamically colder and exhibit 
characteristics (e.g., in S\'ersic indices and velocity dispersions) between 
classical bulges and flattened (oblate) disks. 
Another common feature of pseudo-bulges is their disky shape,
which spurred
\citet[][]{Kormendy1982ApJ...256..460K}
to suggest that secular processes are responsible for their formation. 

One path to improve our understanding of the formation of the two bulge types would
be to quantify the relative importance of different channels of galaxy evolution, 
such as merger-driven versus secular processes \citep[][]{Kormendy04}. 
Recently, instabilities in disks 
\cite[e.g.,][]{Krumholz2010ApJ...724..895K,Bournaud2011ApJ...741L..33B,
Forbes2012ApJ...754...48F,Cacciato2012MNRAS.421..818C}
and mass transfer from unstable disks have been proposed as effective mechanisms to form  
bulges within disk galaxies at high redshifts
\cite[e.g.,][]{Noguchi1999ApJ...514...77N,Elmegreen2008ApJ...688...67E,
Dekel2009ApJ...703..785D,Hathi2009,Genzel2011ApJ...733..101G,Forbes2014MNRAS.438.1552F}. 
Even more recently, \citet[][]{Tonini2016MNRAS.459.4109T}
proposed two distinct populations of bulges: merger-driven bulges,
akin to classical bulges, and instability-driven bulges,
akin to pseudo-bulges. 
\citet[][]{Huertas-Company2015ApJ...809...95H}
also proposed two distinct channels for the growth of bulges in massive galaxies. 
One channel formed around one third (1/3) of the bulges at early epochs (before $z\sim 2.5$) 
through gas-rich mergers or violent disk instabilities — these usually have 
high S\'ersic indices ($n > 3-4$) and small effective-radii ($\rm \sim 1 kpc$). 
The remaining two thirds (2/3) underwent a gradual transformation in morphology at 
late epochs, from clumpy disks to more-regular, 
bulge+disk systems.  Such changes result in significant growth of 
bulges with low S\'ersic indices ($n<3$). 
If such secular evolution is a more important process in forming the bulge population 
at late epochs, bulges should be observed to grow in parallel with disk growth.  

Besides having a disk and bulge, fairly-large and massive spiral galaxies, such as the Milky Way (MW),  often also have extended stellar halos. 
The stellar halo of the Milky Way has been well-characterized (see a review by 
\citealt[][]{Helmi2008A&ARv..15..145H}). 
Recent observational advances have also enabled 
the detection of faint stellar halos around external galaxies, such as M31
\cite[][]{Ferguson2002AJ....124.1452F,Guhathakurta2005AAS...20713501G,
Irwin2005ApJ...628L.105I,Ibata2007ApJ...671.1591I} 
and other nearby disk galaxies
\cite[][]{Mouhcine2005ApJ...633..821M,Mouhcine2005ApJ...633..828M,
deJong2007IAUS..241..503D,Ibata2009MNRAS.395..126I,Mouhcine2010ApJ...714L..12M}.
The observed properties of the stellar halos 
in the Milk Way and its neighbouring galaxies are in general agreement with the predictions 
of $\Lambda$CDM hierarchical galaxy formation models 
\cite[][]{Bell2008ApJ...680..295B,Gilbert2009ApJ...705.1275G,
McConnachie2009Natur.461...66M,Starkenburg2009ApJ...698..567S,Mart2010AJ....140..962M}.
Cosmological simulations predict that the amount of stellar mass in these halos 
should be $\rm \sim 10^8-10^9 M_{\odot}$ for MW-like galaxies, 
and that most of the stellar halo mass would be assembled 
before $z\sim 1$ \cite[][for a review]{DeLucia2008MNRAS,
Cooper2010MNRAS.406..744C,DeLucia12} 

The detection of stellar halos in distant 
disk galaxies has, however, been  scarce, thereby stemming progress 
in tracking  the early formation and assembly histories of disk galaxies. 
Only two works have detected stellar halos beyond $z=0.3$ 
\cite[][]{Zibetti2004MNRAS.352L...6Z,Trujillo2013MNRAS.431.1121T}. 
Exploiting deep, high-resolution {\it HST} images 
in the {\it Hubble Ultra Deep Field} \cite[HUDF;][]{Beckwith2006AJ....132.1729B}. 
\citet[][]{Zibetti2004MNRAS.352L...6Z} detected the stellar halo of a disk galaxy 
at $z = 0.32$ and \citet[][]{Trujillo2013MNRAS.431.1121T}
detected the stellar halos 
of two MW-like galaxies at $z\sim 1$.
Thanks to the large sample of galaxies with HST imaging in the Cosmic Assembly Near-Infrared Deep Extragalactic Legacy Survey
\citep[CANDELS,][]{Grogin11,Koekemoer11}, 
we can now detect and study the stellar halos at intermediate to high redshifts 
in a statistical manner, and significantly advance our understanding of 
the assembly histories of disk galaxies. 

In local galaxies, the isophotal shapes of galaxies are found to be coupled with 
their structural and kinematic properties. 
The isophotes of spheroids often deviate from pure ellipses. 
These deviations originate from the characteristics of 
the stellar orbits that make up these galaxies. The correlations between 
the isophotal deviations and physical properties of galaxies were shown mostly 
for nearby early-type galaxies \cite[e.g.][]{Carter1978,Lauer1985MNRAS.216..429L,Bender1988,Bender1989,hao2006} 
and for a few late-type galaxies \cite[e.g.][]{Erwin2013}. 

Besides isophotal deviations, 
the ellipticity (1 - axis ratio) of a galaxy has been shown to be closely linked 
with the relative importance of ordered rotation and random motion in spheroids \cite[][]{Kormendy13}. 
Furthermore, the ellipticity is linked to the isophotal deviations themselves \cite[][]{hao2006}. 
Generally, more flattened systems (with larger ellipticities) tend to be more rotationally supported 
and have more disky isophotal shapes. Therefore, measurements of radial profiles of 
isophotal ellipticity and deviations are likely to be helpful diagnostics of the kinematics 
and morphological compositions of galaxies. For example, disk galaxies seen nearly edge-on 
appear flattened and have more disky isophotal shapes 
than galaxies dominated by central bulges or outer stellar halos \cite[][]{Zheng15}.    
%
 
Furthermore, quantifying the variation of isophotal shape profiles across cosmic time 
may provide key insights on the evolution of galaxy structure. 
With this motivation, we used the deep, high-resolution {\it HST}/WFC3 imaging data 
to measure the radial profiles of isophotal ellipticity ($\varepsilon$) and 
deviation parameter A$_4$ (defined in \S\ref{measure_s}) for $\sim4,600$ $UVJ$-defined 
SFGs at $0.5<z<1.8$ selected from the CANDELS/GOODS-S and UDS fields. 
The isophotal-shape profiles are well measured to large radii of about three times the semi-major axes for more than 2/3 of 
the galaxies in our sample. For the first time, statistically-robust profile analyses of the isophotal shapes out to large radii 
in distant star-forming galaxies is possible. We study the stacked (median) 
$\varepsilon$ and A$_4$ profiles of our galaxies sub-divided by stellar mass and redshift. 
We also divide our galaxies into \emph{Small} and \emph{Large} SFGs 
which are smaller and larger than the average size-mass relation, respectively. 
The two classes of SFGs 
are found to exhibit statistically significant differences in their radial profiles of the isophotal shape parameters.


The outline of this paper is as follows. Section 2 describes the data 
and the sample selection. Sections 3 details the measurements of isophotal shape profiles. 
We present our main results in Section 4 and finish with a discussion and summary in Section 5. 
Throughout the paper, we adopt a cosmology with $\Omega_{M}$=0.3, 
$\Omega_{\Lambda}$=0.7 and $\rm H=70~km~s$$^{-1}$~Mpc$^{-1}$. 
All magnitudes are in the AB system.

%
%
%
\section{Data and Sample Selection}

\subsection{Data}

The sample of galaxies used in this work is selected from 
the CANDELS/GOODS-S and UDS fields \citep[][]{Grogin11,Koekemoer11}. 
Multi-wavelength photometry catalogs of the two publicly-available fields were built by 
 \citet[][for GOODS-S]{Guo13} and 
\citet[][for UDS]{Galametz13}; they provide details on source identification 
and photometry. Key points for both fields include source detection from the CANDELS 
mosaics in the F160W band with total fluxes of the sources in the {\it HST} bands 
being measured by running {\tt SExtractor} \citep[][]{Bertin96} in dual mode on 
the point spread function (PSF)-matched images. 
Photometry in the lower-resolution images (e.g., ground-based and IRAC) 
was measured using {\tt TFIT} \citep[][]{Laidler07}. 

Redshifts used in this study are spectroscopic, if available, and are otherwise 
photometric redshifts. Photometric redshifts were computed using the offical multi-wavelength 
photometry catalogs described above and adopting a hierarchical Bayesian 
approach. The typical scatter of photometric redshifts spans 
from 0.03 to 0.06  in z \citep[see][for details]{Dahlen13}.
To compute rest-frame total magnitudes from FUV to $K$ band, the redshifts 
are input  to the {\tt EAZY} software package \citep[][]{Brammer2008}, which fits a set of galaxy
spectral energy distribution (SED) templates to the multi-wavelength photometry. 
For stellar masses, we adopt the CANDELS official values 
released by \citet[][]{Santini2015}, which are the median of 
ten separate SED fitting results \citep[][]{Mobasher2015} after 
being scaled to the \citet[][]{Chabrier03} initial mass function (IMF). 
The typical formal uncertainty of stellar masses is 
$\sim 0.1$ dex \citep[see][for details]{Santini2015}.
 
Spatially-resolved photometry is taken from the {\it HST}-based 
multi-band and multi-aperture photometry catalogs of CANDELS still 
under construction by Liu et al. (in preparation). These catalogs include 
the radial profiles of isophotal ellipticity ($\varepsilon$) and 
disky/boxy parameter $\rm A_4$ in both F125W(J) and F160W(H), and 
the observed surface brightness profiles in all {\it HST}/WFC3 and 
ACS bands if available. The detailed procedure of isophotal measurement 
is presented in \S3.


Global-galaxy structural parameters measured by \cite{vanderWel2012}
with {\tt GALFIT} \citep[][]{Peng02} 
are available for all galaxies in two fields.
Images of each galaxy in both F125W and F160W were fit with a single-S\'ersic model,
yielding the best-fitting S\'ersic index ($n$), effective radius along the semi-major axis ($R\rm _{SMA}$), 
axis ratio (b/a), and position angle (PA), along with estimates of their errors.
We use $R\rm _{SMA}$ as our indicator of galaxy size, rather than
circularized effective radius, $R\rm _{eff}$, because the latter depends 
on the axis ratio ($R\rm _{eff} \equiv \sqrt{b/a} \times$$R\rm _{SMA}$), 
and $R\rm _{SMA}$ is a more faithful indicator of the intrinsic size for oblate systems.


\subsection{Sample Selection}

The full GOODS-S and UDS catalogs contain 34,930 \citep[][]{Guo13} 
and 35,932 \citep[][]{Galametz13} objects, respectively. 
The parent sample used in our analysis is constructed 
by applying the following criteria to the catalogs: \\

1. Observed F160W($\rm H$) magnitude brighter than 
24.5 and
the {\tt GALFIT} quality $\rm flag = 0$
in F125W for $z<1$ and F160W for $z>1$ \citep[][]{vanderWel2012} 
to ensure well-constrained {\tt GALFIT} measurements 
and to eliminate doubles, mergers, and disturbed objects. 
Table 1 shows that only about a quarter of galaxies in the combined sample 
of GOODS-S and UDS satisfy this criterion. \\ 

2. SExtractor Photometry quality flag $\rm PhotFlag = 0$
to exclude spurious sources; \\

3. SExtractor $\tt CLASS\_STAR$ $< 0.9$ to reduce contamination by stars; \\

4. Redshifts between $0.5<z<1.8$ and stellar masses of
$\rm 9.0<logM_{\ast}/M_{\odot}<11.0$ to maintain a high 
mass-completeness limit for SFGs ($\sim100\%$ at $z=0.5$ and $\sim85\%$ at $z=1.8$) 
\citep[][]{vandelerl14} and 
to assure that all isophotal parameters can be measured in similar, rest-frame optical bands. \\

5. Well-constrained measurements of isophotal parameters
(Isophotal $\rm PhotFlag = 0$) from Liu et al. (in preparation); \\ 

6. $\rm R_{SMA} > 0.18\arcsec$ (3 drizzled pixels) to reduce the
effect of PSF smearing on isophotal measurements; \\

7. SFGs are selected from rest-frame $UVJ$ diagrams by ($(U-V) < 0.88 \times (V-J)+0.49$ for $z>1$ and
$(U-V) < 0.88 \times (V-J)+0.59$ for $z<1$) following the criteria defined by \citet[][]{williams09};\\

8. Exclude a few compact SFGs (cSFGs) with the criterion 
of $\rm log\Sigma_{1.5}>$10.3 from \citet[][]{Barro2013}, 
since these cSFGs might start as compact quiescent galaxies 
at high redshifts and later evolve into larger quiescent galaxies 
at lower redshifts \citep[e.g.,][]{Barro2013,Barro2017}. \\

After the above cuts, 4,595 SFGs remain: 2,033 from GOODS-S and 2,562 from UDS. 
Table 1 lists the resulting sample size after applying each selection criterion. 
The top panels of Figure 1 present the rest-frame $UVJ$ diagrams for 
our galaxies in three redshift bins. 
SFGs defined by the criteria of \citet[][]{williams09} are shown with blue dots 
and color-coded by $\rm logR_{SMA}$. 
Quiescent galaxies are excluded in this work and they are located in the gray hatched upper 
corners of the diagrams. 
The bottom panels of Figure 1 show the size($\rm logR_{SMA}$)-mass relations for $UVJ$-defined 
SFGs in the three adopted redshift bins. The size-mass panels are color-coded by 
the global ellipticity defined as $\rm \varepsilon_{global} = 1 - (b/a)_{Galfit}$.
To derive the mean size-mass relations, an initial fit to all SFGs is made; objects more 
than 2$\sigma$ away from the fit are then excluded for the next fit. This fitting process is repeated 
until no new objects are excluded. The parameters of the final fits are presented in Table 2. 
The solid black lines in the bottom panels of Figure 1 indicate the adopted 
best-fit linear relations to the galaxies. 
After the fits are done, vertical offsets in $\rm logR_{SMA}$ from the relations are calculated 
for our SFGs in each bin. The offset for a given galaxy is denoted by $\rm \Delta log R_{SMA}$.
For simplicity, we hereafter refer to galaxies (in a given mass and redshift bin) 
with $\rm \Delta log R_{SMA} >0$ (i.e., larger than average) as \emph{Large} SFGs (LSFGs), 
and galaxies with $\rm \Delta log R_{SMA} <0$ as \emph{Small} SFGs (SSFGs).
Note that our slopes are systematically shallower 
by $\sim 0.1$ dex compared to the fits by \citet[][]{vandelerl14}, 
probably due to the exclusion of very small galaxies with $\rm R_{SMA} < 0.18\arcsec$.   
These small discrepancies do not affect our results, since we are only concerned with 
relative-size differences at fixed mass and redshift. 

To examine evolutionary trends as a function of both redshift and mass at the same time, 
we divide the sample into four mass bins ($\rm 9.0<logM_*<9.5$, $\rm 9.5\leq logM_*<10.0$,
$\rm 10.0\leq logM_*<10.5$ and $\rm 10.5\leq logM_*<11.0$)
and three redshift bins ($0.5<z<1.0$, $1.0<z<1.4$, and $1.4<z<1.8$).
This 4x3 grid of diagrams is a powerful visualization tool
to track the movement of galaxies as they evolve in mass with time.  
Figure 2 shows the distributions of $\rm \varepsilon_{global}$ and corresponding
median values ($\rm \varepsilon_{global,med}$) for SSFGs and LSFGs in each mass-redshift bin, respectively.
To recognize the intrinsic structure of galaxies more easily, 
we further divide our sample galaxies  
into two sub-classes: ``edge-on'' ($\rm \varepsilon_{global}> \varepsilon_{global,med}$) 
and ``face-on'' ($\rm \varepsilon_{global}<\varepsilon_{global,med}$), 
according to the relative observed ``inclination''. 
So that our isophotal analyses are done in similar rest-frame optical bands, 
we measure the isophotal profiles in F160W band for $z>1$ galaxies and 
in F125W band for $z<1$ galaxies.


\section{Measurement of Isophotal Shape Profiles}\label{measure_s}

The radial profiles of galaxy isophotal parameters, $\varepsilon$ and $\rm A_4$, 
used in this work come from the {\it HST}-based, multi-wavelength 
and multi-aperture photometry catalogs built by Liu et al. (in preparation). 
The isophotal parameters are measured by using the {\tt IRAF} routine {\tt ellipse} 
within STSDAS, which is based on a technique described by \cite{Jedrzejewski1987}. 
We now summarize the measurement process. 

First, we trim the original PSF-matched mosaic images in each band 
to generate multi-band cut-out images centered on each target galaxy.
Before the ellipse fitting, {\sc SExtractor} is used to identify sources 
within the detection limit; they are removed then to obtain a background-only image for each band. 
Median filtering is applied to derive a local-background image in each band, 
using sigma clipping to reduce the impact of relatively-brighter, residual background-sources. 
This local background is then subtracted from each cut-out image.
We verify that the flux distributions in empty regions of each cut-out image 
in each band are centered on zero after the background subtraction. 

After background subtraction, we run {\sc SExtractor} again 
on the trimmed image to generate a {\sc `SEGMENTATION'} image, 
which identifies all objects with flags in the image. 
A mask image, with all the detected objects flagged except the
galaxy of interest, is then obtained from the {\sc
`SEGMENTATION'} image. We carefully examine all the mask images 
in each band and correct a few bad images manually to create 
good mask images for all galaxies. Photometry is then 
performed on the trimmed images with the masked areas excluded from the reduction.

We use the geometric center, ellipticity, position angle and 
effective radius along the semi-major axis of sample galaxies 
obtained from the {\tt GALFIT} measurements by \cite{vanderWel2012} 
as initial values in the ellipse fitting.
In the \texttt{ellipse} task, the image intensity is first sampled along a 
trial ellipse generated using these parameters, and the intensity string
$I(\theta)$ is expanded in a Fourier series, 
\begin{equation}
I (\theta) = I_0 + \sum_{n=1}^N [ A_n \cos (n \theta) + B_n \sin (n \theta) ]
\label{gs}
\end{equation}
where $I(\theta)$ is the intensity (in units of ADU~s$^{-1}$~pixel$^{-1}$)
on the ellipse in the direction of $\theta$,
$I_{0}$ is the average intensity of the ellipse,
the position angle $\theta$ is defined to be 0\arcdeg\ along the positive
$y$-axis and increases counter clockwise.
$N$ is the highest harmonic fitted,
$A_n$ and $B_n$ are the Fourier coefficients. 
The most significant non-zero component of the 
Fourier analysis is the $A_4$ parameter (corresponding to the 
$cos(4\theta)$ term). Using the sign of this parameter, the isophote of 
a galaxy is classified as disky ($A_4>0$) or 
boxy ($A_4<0$).

During the fitting, we allow the geometric center, ellipticity and position 
angle to vary freely. Successive ellipses are fitted along the major axis, 
starting from the effective radius and moving inward and outward 
with logarithmic steps of 0.3, until the ellipse fitting process fails to converge.
The output of {\tt ellipse} is a table containing the radial profiles of 
several isophotal parameters (along with their uncertainties), 
such as ellipticity ($\varepsilon$), Fourier coefficients (i.e., $\rm A_4$), 
position angle (PA), and surface brightness in each elliptical annulus.
Figure 3 illustrates our measurements for three 
nearly edge-on galaxies 
(GOODS-S~14994, GOODS-S~22208 and GOODS-S~19762) and Figure 4 for
three nearly face-on galaxies (GOODS-S~10421, GOODS-S~26255 and UDS12524) 
in three different redshift bins. 

We stress the importance of two steps in the above process: the local 
background subtraction and the use of large logarithmic steps.
Both significantly improve the accuracy of the measurements.
More technical details  will be included
in the documentation of the Liu et al. (in preparation) catalogs.


\section{Results}

To derive the stacked $\varepsilon$ and A$_4$ profiles 
for each sub-class in every mass-redshift bin, we adopt {\tt IRAF}/{\tt proto} to do cubic spline interpolation
and then compute the median value and 68\% distribution of the scatter for 
every selected position in each bin.  
The resulting median $\varepsilon$ and A$_4$ as a function of 
normalized radius ($\rm \widetilde{R}=R/R_{SMA}$) for SSFGs and LSFGs are shown in Figures 5 to 8. 
Note that not every galaxy has a reliable measurement of isophotal profile that extends to large radii. 
The median values at the positions where the fractions of 
accurate data points are below $1\sigma$ are removed in our analysis.

To quantify the general trends of composite $\varepsilon$ and A$_4$ profiles in 
each mass-redshift bin, we use the {\tt segmented} package in {\tt R} programming language 
to fit the two-broken-lines model to each profile outside the PSF FWHM (0.18$\arcsec$). 
The model is derived simultaneously yielding point estimates and their 
approximate standard errors for all the model parameters, 
including the break-point where the linear relation changes.
If the two broken lines model fails to fit a profile, the program automatically 
generates a single linear regression model. 
The best-fit parameters are given in Table 3, including the 
slopes, intercepts, relevant standard errors, break-radii, and 
R-squared values. The R-squared value quantifies how
well the model fits the data. R-squared values close to 1 indicate excellent fits while those close to zero indicate poor fits. 
Table 3 shows that the majority of segmented models 
are good enough to trace the trends of our profiles.
The best models with $2\sigma$ lower and upper limits are shown 
shaded in Figures 5 to 8.

Next, we summarize the observed trends of $\varepsilon$ and A$_4$ profiles with mass and viewing angle. 
First, we discuss the trends observed in low-mass galaxies before those in high-mass galaxies.

1. At low masses ($M_{\ast} < 10^{10}M_{\odot}$), the SSFGs ($\rm \Delta log R_{SMA} <0$) 
generally have nearly flat $\varepsilon$ profiles (the majority of segmented models have 
slope values within $\pm0.2$ of zero) 
in both face-on and edge-on views, especially at $z>1$ (Figure 5). 
The average ellipticity values are $\sim0.4$ for edge-on and $\sim0.2$ for face-on, which 
imply that these systems are not intrinsically highly flattened. 
The flat $\varepsilon$ profiles in face-on views are generally consistent 
with our expectations and, therefore, are not surprising. 
In contrast, if these galaxies actually harbor disks and are being viewed edge-on, 
a significant change in $\rm \varepsilon$ with increasing radius should be observed \citep[][]{Zheng15}. 
Such trends, however, are not observed in edge-on views. 
These findings indicate that these galaxies are likely composed of a single structure and it is not disk-like.
In addition, we find that the SSFGs in these bins also have nearly flat A$_4$ profiles 
(the slopes of segmented models are within $\pm0.02$ of zero) 
in both face-on and edge-on (Figure 6); furthermore, the median values of A$_4$ at all radii 
are almost zero. These findings imply that, statistically, 
these galaxies are not rotationally supported and, thus, they are not disks. 
These systems may not be spheroidal either, because 
they usually have S\'ersic indices $n\sim1$ \citep[][]{Wuyts11}. 
\citet[][]{hao2006} showed the relation between $\rm \varepsilon$ and 
A$_4$ for nearby massive spheroids and find that spheroidal systems 
with $\rm \varepsilon=0.4$ are usually disky (A$_4>0$). If this relation were to hold 
for low-mass systems at higher redshifts, it would provide another argument against our
observed sample being spheroids.  
Furthermore, it can be seen in Figure 6 that the A$_4$ points of these galaxies distribute 
on both sides of the A$_4=0$ lines randomly, and the face-on and edge-on systems 
are mixed together. The findings suggest that these galaxies are likely to have 
irregular structures without distinct boundaries of regular components. 

Likewise, the low-mass ($M_{\ast} < 10^{10}M_{\odot}$) LSFGs ($\rm \Delta log R_{SMA} >0$) 
in the same bins also have relatively flat $\varepsilon$ profiles in face-on views (Figure 7). 
When seen in edge-on, however, these large systems have $\varepsilon$ profiles 
that mainly increase monotonically with radius (the slopes of inner fits are 
greater than $\sim0.2$). Although some LSFGs 
have decreasing $\varepsilon$ profiles in their outermost regions, these trends 
are relatively weak compared to those observed in high-mass galaxies.

For the low-mass LSFGs, the average ellipticity values are $\sim0.55$ 
for edge-on views and $\sim0.35$ for face-on views. 
This result implies that these systems are more flattened intrinsically 
than the SSFGs in the same mass-redshift bins. 
The low-mass LSFGs exhibit more positive 
A$_4$ (disky) profiles that dominate in the intermediate regions 
when seen edge-on than face-on (Figure 8). 
This finding implies that these large galaxies 
likely have disk-like components flattened by rotation.     

2. At high masses ($M_{\ast} > 10^{10}M_{\odot}$), the SSFGs also have 
nearly flat $\varepsilon$ profiles in face-on views (Figure 5). 
In edge-on views, however, the $\varepsilon$ profiles of SSFGs exhibit a 
distinctly different pattern compared to those for low-mass SSFGs: significant increase 
with radius in the inner regions 
(the slopes  are
greater than $0.2$); goes through a maximum at intermediate radii; followed by decrease 
in the outskirts (the slopes become negative). 
Such trends are more prevalent for more massive ($M_{\ast} > 10^{10.5}M_{\odot}$) galaxies 
or at lower redshifts ($z<1.4$) and  especially clear for the LSFGs (see Figure 7). 
Even for face-on views, similar trends are seen 
for the most massive ($M_{\ast} > 10^{10.5}M_{\odot}$) LSFGs. 
The trend is likely due to the presence of three components: 
bulges in the inner regions, disks in the intermediate regions, and halo-like stellar components in the outskirts. 
When viewed edge-on, galaxies should exhibit ellipticity profiles that reveal
the relative flattening of various components. 
The central bulge and outer stellar halo will appear much rounder than 
the disk in the intermediate region.
Similar trends can be found  in the corresponding 
A$_4$ profiles of edge-on systems (Figures 6 and 8): the isophotes 
in the intermediate regions are obviously disky with positive A$_4$, whereas the inner and outer isophotes 
are close to perfect ellipses (A$_4\sim0$). 
Furthermore, compared to the low-mass ($M_{\ast} < 10^{10}M_{\odot}$) galaxies 
at the same redshifts, the intermediate regions of high-mass galaxies are more disky (with larger A$_4$ values) 
in edge-on views. This result implies that these massive galaxies likely 
possess disks with relatively faster rotation.

To clearly trace the variation of ellipticities in the inner regions 
as galaxies evolve in mass with time, we present in Figure 9 the distributions of ellipticity 
at $\rm R=1.5 kpc$ ($\varepsilon_{1.5}$) for the SSFGs and LSFGs in each mass-redshift bin. 
This physical radius ($\rm R=1.5 kpc$) is relatively close to the centers of galaxies 
but the ellipticity measurement at this radius is less affected by PSF smoothing 
compared to the very center at $\rm R_{SMA} < 0.18\arcsec$, as indicated in Figures 5 and 7. 
One sees that, as galaxies evolve 
towards high stellar masses and low redshifts, the $\varepsilon_{1.5}$ of 
both SSFGs and LSSFGs tend to decrease (become rounder) statistically. 

To double-check our results, we visually inspect the images of 
sample galaxies carefully. In Figure 10, we present cut-out images of 4 examples of highly-inclined  systems 
for different mass-redshift bins. In each panel, the upper two images are for LSFGs 
and the lower two are for SSFGs. 
Obviously, in the low-mass ($M_{\ast} < 10^{10}M_{\odot}$) and high-redshift ($z>1$) bins, 
the images of SSFGs exhibit good consistency with no obvious disk-like features (mostly irregular morphology). 
In contrast, the images of LSFGs in the same bins indeed exhibit possible 
disk-like structure. 
As galaxies evolve towards the high masses ($M_{\ast} > \sim10^{10}M_{\odot}$) 
and low redshifts ($z<\sim1.4$), the disk component becomes more prominent in both LSFGs and SSFGs. 
Meanwhile, the diffuse, halo-like stellar components appear to dominate the outer regions of galaxies.
The visual inspection is in good agreement with our quantitative analysis of the isophotal structure. 


\section{Discussion and Summary}

We have measured the radial profiles of isophotal ellipticity ($\varepsilon$) and 
A$_4$ out to radii of $\sim3R_{\rm SMA}$ in similar, rest-frame, optical bands 
for $\sim$4,600 SFGs between redshift 0.5 and 1.8 in the CANDELS/GOODS-S and UDS fields. 
With this sample, we study the stacked $\varepsilon$ and A$_4$ profiles 
on an evolutionary grid laid out by stellar mass and redshift \citep[see Figure 6 in][]{Fang2017} .
The grid of sub-panels is a useful visualization tool to track the movement of galaxies 
as they evolve in stellar mass \citep[][]{Moster2013,Papovich2015}.
For the first time, we are able to undertake a statistically robust analysis of the isophotal structure of 
distant star-forming galaxies. 
The mean size-mass relation in a given redshift bin is used as the divider between   
the relatively \emph{Small} SFGs (SSFGs) and \emph{Large} SFGs (LSFGs). 
We find that, statistically, these two classes exhibit different radial patterns of 
isophotal shape. 
Our main conclusions are as follows:

1. At low masses ($M_{\ast} < 10^{10}M_{\odot}$), the SSFGs generally have nearly flat $\varepsilon$ and A$_4$ profiles 
in both edge-on and face-on views, especially at $z>1$. 
Moreover, the median A$_4$ values at all radii are almost zero with no disky or boxy signatures,  
but the A$_4$ distributions have quite large scatter. 
In contrast, the more-inclined (edge-on) LSFGs in the same mass-redshift bins 
generally have $\varepsilon$ profiles that mainly increase monotonically 
with radius and disky profiles (A$_4>0$) that dominate 
in the intermediate regions. 
The findings imply that SSFGs are not disk-like, whereas LSFGs 
likely have disk-like components flattened by significant rotation. 
  
2. At high masses ($M_{\ast} > 10^{10}M_{\odot}$), both more-inclined SSFGs and LSFGs 
generally exhibit distinct $\varepsilon$ and A$_4$ profiles  
that first increase with radius, then reach maxima, and finally decrease. 
Such trends are more prevalent at lower redshifts ($z<1.4$) 
or for more massive ($M_{\ast} > 10^{10.5}M_{\odot}$) galaxies. 
This profile pattern can be explained by the galaxies possessing central bulges,
disks in the intermediate region,  and halo-like stellar components in the outskirts.
Compared to the low-mass SFGs at the same redshifts, the intermediate isophotes of edge-on massive SFGs are more disky, 
indicating that these massive galaxies likely have disks with faster rotation.  

3. Central ellipticities of both SSFGs and LSFGs tend to decrease (become rounder) with increasing mass 
and decreasing redshift. Moreover, the peak values of both ellipticity and A4 tend to 
increase as galaxies increase in mass with time. These findings suggest bulges enlarge 
with disk growth.

Recently, \citet[][]{vanderwel2014} derived the intrinsic, 3-dimensional distributions of 
global axis ratio ($b/a=1-\varepsilon$) of distant SFGs through the observed ellipticity distribution. 
They showed that the low-mass ($M_{\ast} < 10^{10}M_{\odot}$) SFGs at $z>1$ 
possess a broad range of geometric shapes, and the fraction of non-disk (probably prolate) galaxies 
increases at higher redshifts and lower masses. 
This result is consistent with our finding. 
More recently, \citet[][]{liu2016} showed that the rest-frame $NUV$-$B$ color gradients 
in low-mass ($M_{\ast} < 10^{10}M_{\odot}$) SFGs at $z\sim1$ are generally flat 
after correcting for dust reddening, which implies that the newly-formed stars in these galaxies 
may be randomly mixed with older populations. These galaxies may be supported predominantly 
by random motions, a result also consistent with our A$_4$ analysis. 
Recent cosmological hydrodynamical zoom-in simulations by \citet[][]{Ceverino14} show that 
low-mass galaxies at high redshifts are sometimes elongated, bar-like systems with irregular morphology. 
The large and homogeneous survey with integral field spectroscopy (IFS) 
by KMOS \citep[e.g.,][]{Wisnioski2015} reveals that $\sim83\%$ of SFGs with 
$M_{\ast}=3\times 10^{9} - 7\times10^{11}~M_{\odot}$ at $0.7<z<2.7$ are 
rotation dominated and $\sim70\%$ of SFGs are disk-like systems. 
Galaxies that are resolved by KMOS, but not rotating, are found primarily at low stellar 
masses. These findings are also qualitatively consistent with our isophotal analyses. 

We caution that whether the change in the ellipticity 
is observed for single-component rotating disks is unclear. Also, the relation 
between positive A$_4$  and fast rotation is seen only in nearby 
early-type galaxies \cite[e.g.,][]{hao2006}. 
Whether single-component disk galaxies tend to have positive A$_4$ values is also not clear.
We interpret the increasing inner and decreasing outer profiles of ellipticity and A$_4$ 
as the consequence of galaxies having multiple components of inner bulges, intermediate-scale disks, and outer stellar halos. 
It should be sobering that star-forming galaxies often also have  additional substructures,  
such as spiral arms, rings, clumps, and bars \cite[e.g.][]{Erwin2002,Guo2015,Cheung2013}. 
The spiral arms, clumps, and rings tend to appear randomly and they may not affect 
our profiles once they are mixed in our stacks, whereas bars could have significant effects on our results \cite[e.g.][]{Erwin2013}.

We show the distributions of S\'ersic index ($n$) of our sample galaxies in the same mass-redshift grid 
in Figure 11. Both LSFGs and SSFGs exhibit increasing S\'ersic indices
as galaxies evolve towards the high masses and low redshifts. 
This trend indicates that the fraction of bulge-dominated systems likely increases as galaxies evolve,
a result in agreement with previous studies
\cite[e.g.][]{vanderWel2011ApJ...730...38V,Bruce2012MNRAS.427.1666B,
Huertas-Company2015ApJ...809...95H,
Margalef-Bentabol2016MNRAS.461.2728M,Barro2017}. 
Note that Sersic indices here are not those of bulges but, instead, are those of the entire galaxy. 
Investigations of  bulge Sersic indices will be possible with the advent of improved spatial resolution and higher S/N provided by {\it JWST}.

What leads to decreasing  ellipticities in the outskirts of the highly-inclined massive galaxies 
is quite intriguing. 
Based on our isophotal analyses and visual inspections of images,
we propose halo-like stellar 
components in the outer region. 
\cite{Zheng15} showed that the composite ellipticity profile of nearby disk galaxies rises slowly 
between $\rm \sim0.5 R_{90}$ and $\rm \sim1.4 R_{90}$ and then slowly declines out to 
$\rm \sim3 R_{90}$ (see Figure 11 in their paper). For a pure disk galaxy having 
an exponential radial surface brightness profile, $\rm R_{90}$ is roughly twice the effective 
radius. The transition radii in the ellipticity profiles of our massive SFGs are 
roughly between $\rm R_{SMA}$ and $\rm 2R_{SMA}$, which are slightly smaller than,
but already comparable to, that of local disk galaxies reported by \cite{Zheng15}. 
We also show that the outer ellipticity and A4 beyond $\rm 2R_{SMA}$,
seem to be constant over time, which might favor scenarios that have halo-like stellar components already existing 
at high redshifts (i.e., $z\sim2$). Our finding should be regarded with  caution,  
since cosmological surface-brightness dimming makes observations of 
faint stellar halos at high redshifts extremely difficult.
\cite{Zheng15} further showed that the characteristic radial profiles of 
color, stellar mass-to-luminosity ratio ($M/L$), and stellar age of nearby disk galaxies 
have  a ``U'' shape (they first decline with increasing radius, but then rise in 
the outer region). The minima are also located at radii of around 0.8 to 1.0 $\rm R_{90}$ or 
at locations where the local stellar mass surface density is $\rm \sim10M_{\odot}/pc^{-2}$. 
These findings further support the contribution of halo light beyond such radii. 
With these observational results, \cite{Zheng15} argued that a combination of a radial migration of stars in 
the inner region and a truncation of recent star formation in the outer part is 
likely to be required to regulate the evolution of disk galaxies. 
The general trend for nearby disk galaxies to have redder (older) outer 
disks and stellar halos has also been reported by other studies 
\cite[e.g.][]{Bako2008ApJ,Bakos2012arXiv,Yoachim2010ApJ,Yoachim2012ApJ,
Gonzalez2014A&A}. 
In a follow-up of our current work, we plan to explore their possible progenitors 
at moderate redshift.


\section*{Acknowledgements}

We acknowledge the anonymous referee for a constructive report that significantly 
improved this paper. This project was supported by the NSF grants of China No.11573017 and 11733006. 
We acknowledge support of the CANDELS program HST-GO-12060 by NASA through a grant from 
the Space Telescope Science Institute, which is operated by the Association of Universities 
for Research in Astronomy, Incorporated, under NASA contract NAS5-26555. 
S.M.F., Y.G., D.C.K., and H.M.Y. acknowledge partial support from US NSF grant AST-16-15730. 
X.Z.Z. thanks support from the National Key Research and Development Program of 
China (No. 2017YFA0402703), NSFC (grant 11773076) and the Chinese Academy of Sciences (CAS) 
through a grant to the CAS South America Center for Astronomy (CASSACA) in Santiago, Chile.



\clearpage

\begin{figure*}
\centering
\includegraphics[angle=0,width=\textwidth]{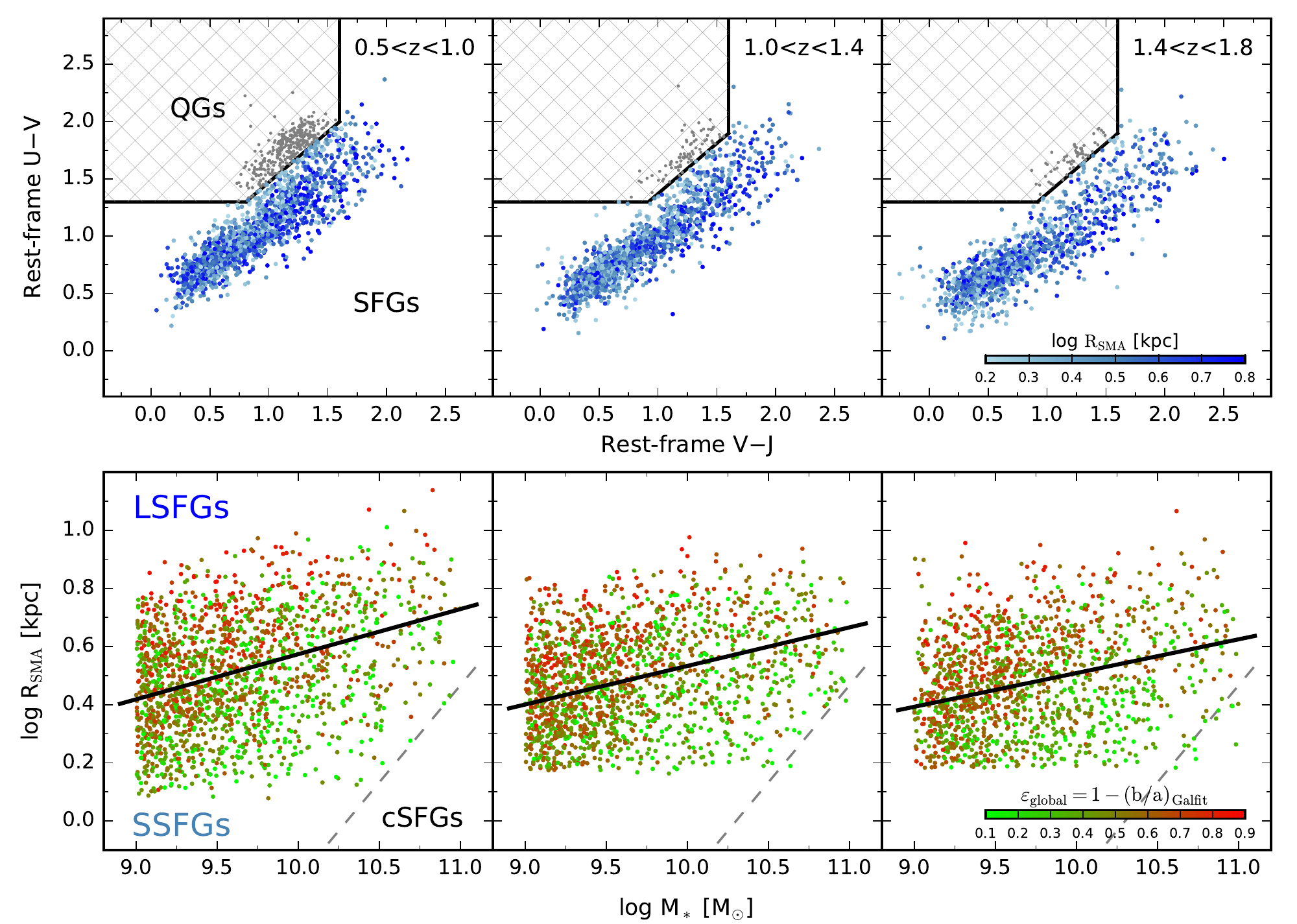}
\caption{
Rest-frame $\rm UVJ$ color-color diagrams (top) for our sample galaxies
after applying the selection cuts
and the size-mass relations (bottom) for $\rm UVJ$-defined SFGs only
in three redshift bins, respectively.
In the top panels, the solid lines are the classification criteria of
\cite{williams09}. SFGs are shown with solid points
and color coded by $\rm log R_{SMA}$. 
Quiescent galaxies are indicated by gray hatching.
In the bottom panels, the solid black lines indicate the best-fit linear 
relations in three redshift bins, respectively.  
The vertical offsets from the relations ($\rm \Delta log R_{SMA}$) 
are used to divide our SFGs into LSFGs ($\rm \Delta log R_{SMA} >0$) 
and SSFGs ($\rm \Delta log R_{SMA} <0$) in each redshift bin.   
The grey dashed lines indicate the classification 
criterion of compact and non-compact SFGs by \citet[][]{Barro2013}.
Data points are color-coded by global ellipticity 
($\rm \varepsilon_{global}=1-(b/a)_{Galfit}$). 
\label{fig1}}
\end{figure*}

\begin{figure*}
\centering
\includegraphics[angle=0,width=\textwidth]{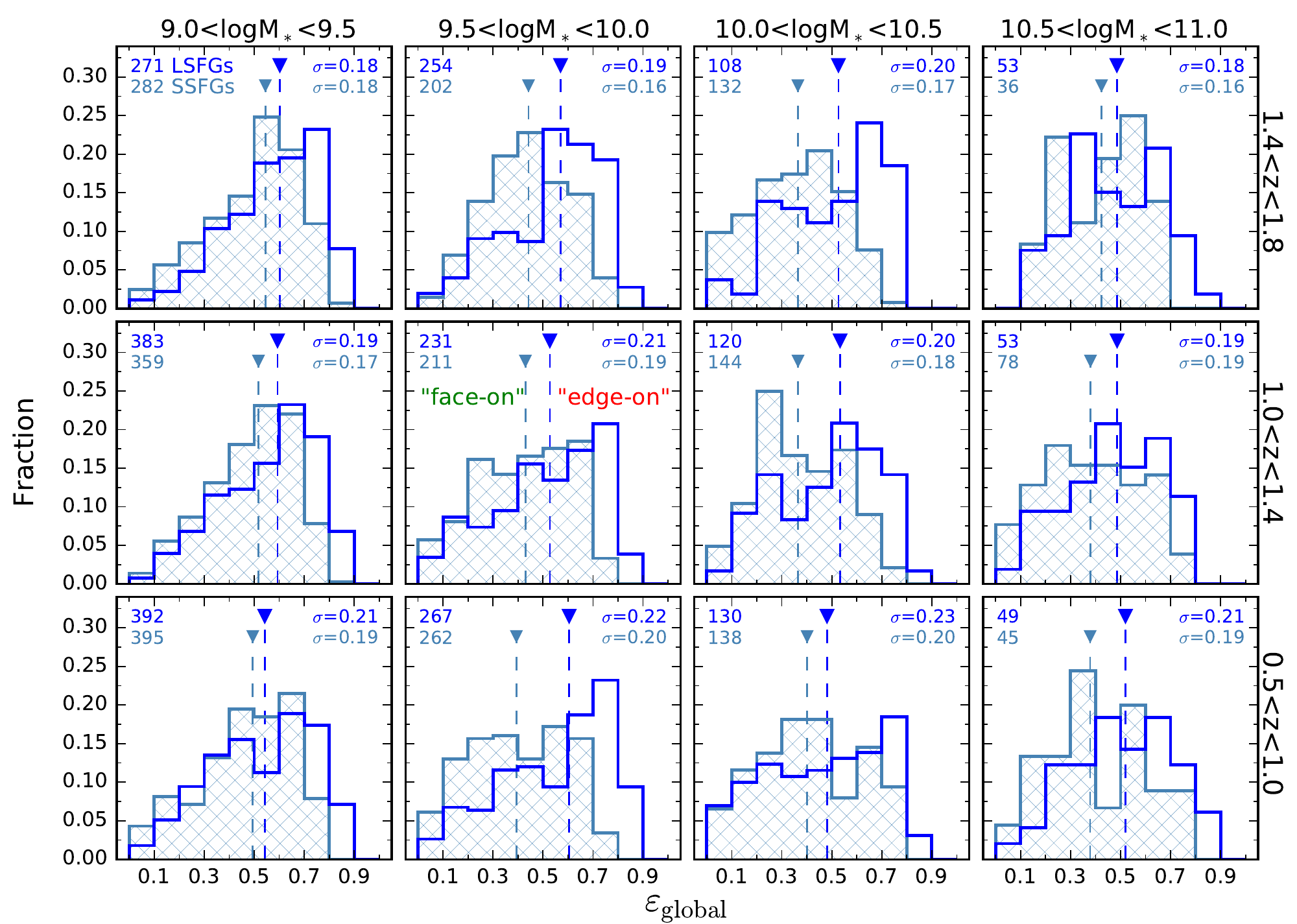}
\caption{
Distributions of global ellipticity for LSFGs (blue) and SSFGs (steel blue) in each
mass-redshift bin, respectively.
The median values are indicated with triangles plus dashed lines. 
The standard deviations ($\sigma$) are presented in the right-top corner of each panel.
The galaxy numbers are presented in the left-top corner of each panel.
\label{fig2}}
\end{figure*}

\begin{figure*}
\centering
\includegraphics[angle=0,width=\textwidth]{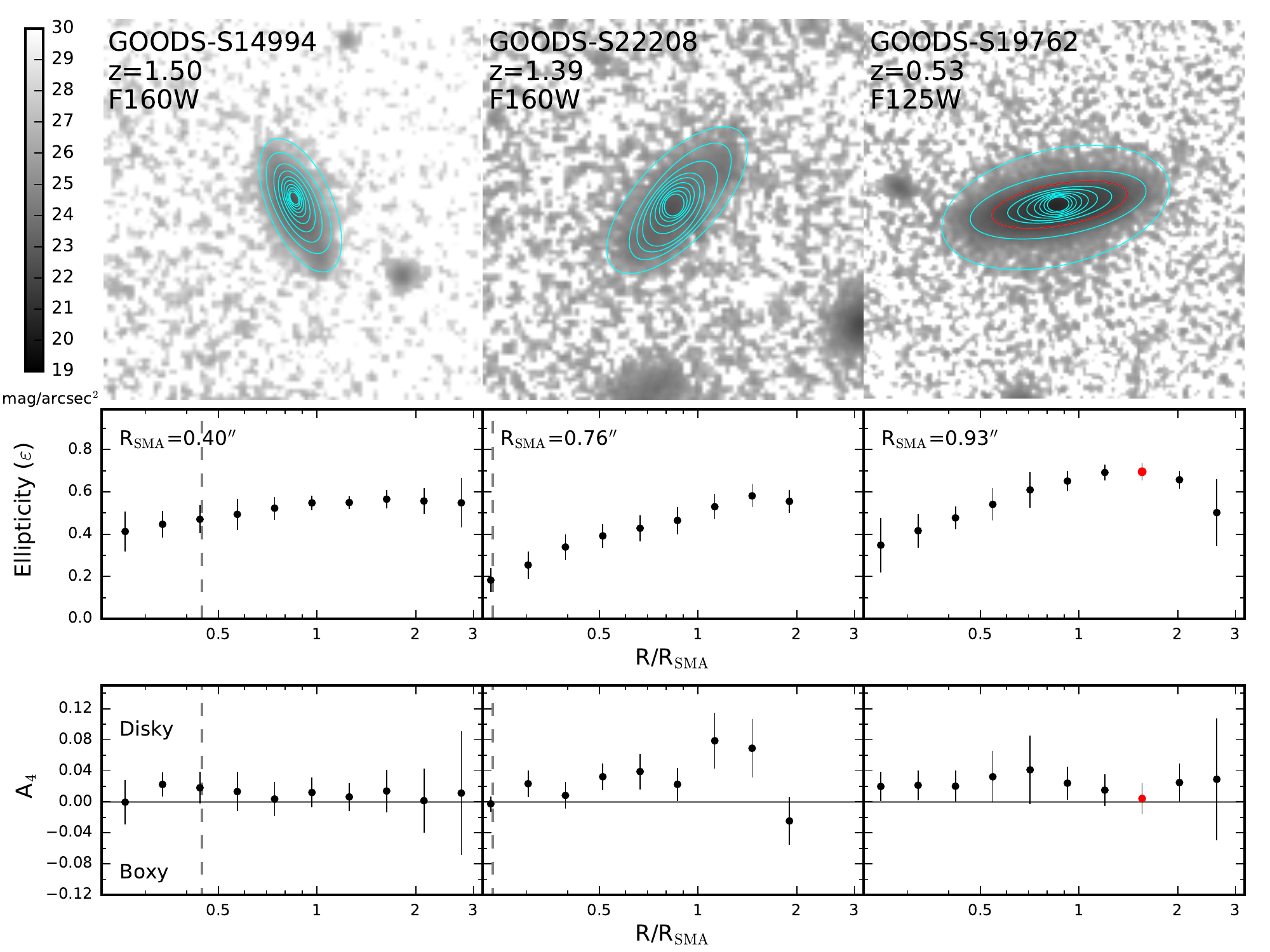}
\caption{
Example nearly edge-on galaxies GOODS-S~14994, GOODS-S~22208 and GOODS-S~19762 illustrating
the measurement of isophotal shape profiles.
In the top panels,
the cyan ellipses are from our isophotal measurements by $\tt IRAF/ellipse$. 
The red ellipse in the right panel indicates the position of the maximum ellipticity. 
The corresponding radial ellipticity profiles with errorbars are shown in the 
middle panels. The corresponding A$_4$ profiles with errorbars are shown in 
the bottom panels. The vertical dashed lines in the middle and bottom panels 
indicate the FWHM of PSF ($0.18{\arcsec}$). The solid horizontal lines 
in the bottom panels indicate A$_4$ = 0.   
\label{fig3}}
\end{figure*}

\begin{figure*}
\centering
\includegraphics[angle=0,width=\textwidth]{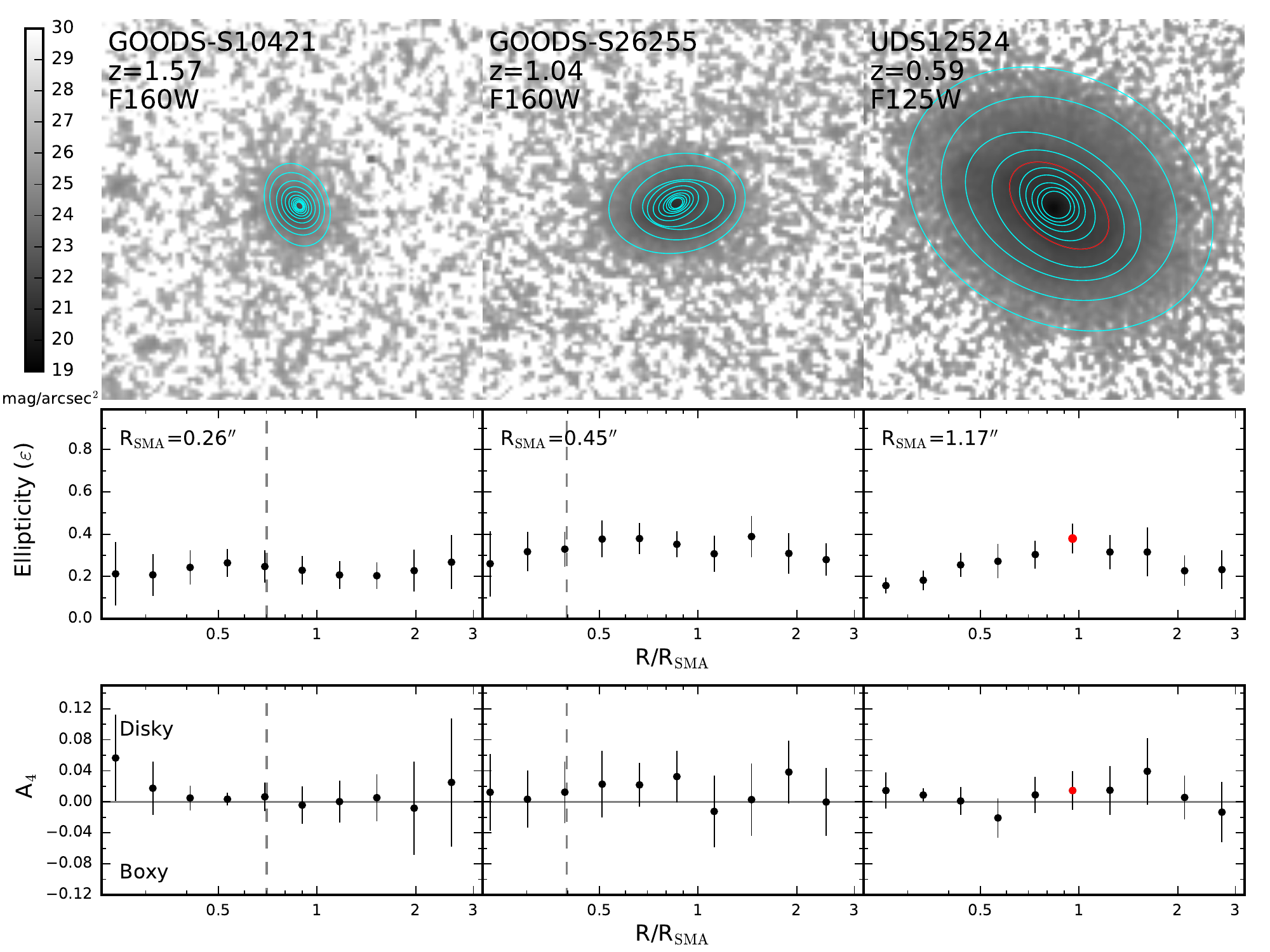}
\caption{
Example nearly face-on galaxies GOODS-S~10421, GOODS-S~26255 and UDS~12524 illustrating
the measurement of isophotal shape profiles.
The data points, error-bars and lines have the same meanings as those in Figure 3.
\label{fig4}}
\end{figure*}

\begin{figure*}
\centering
\includegraphics[angle=0,width=\textwidth]{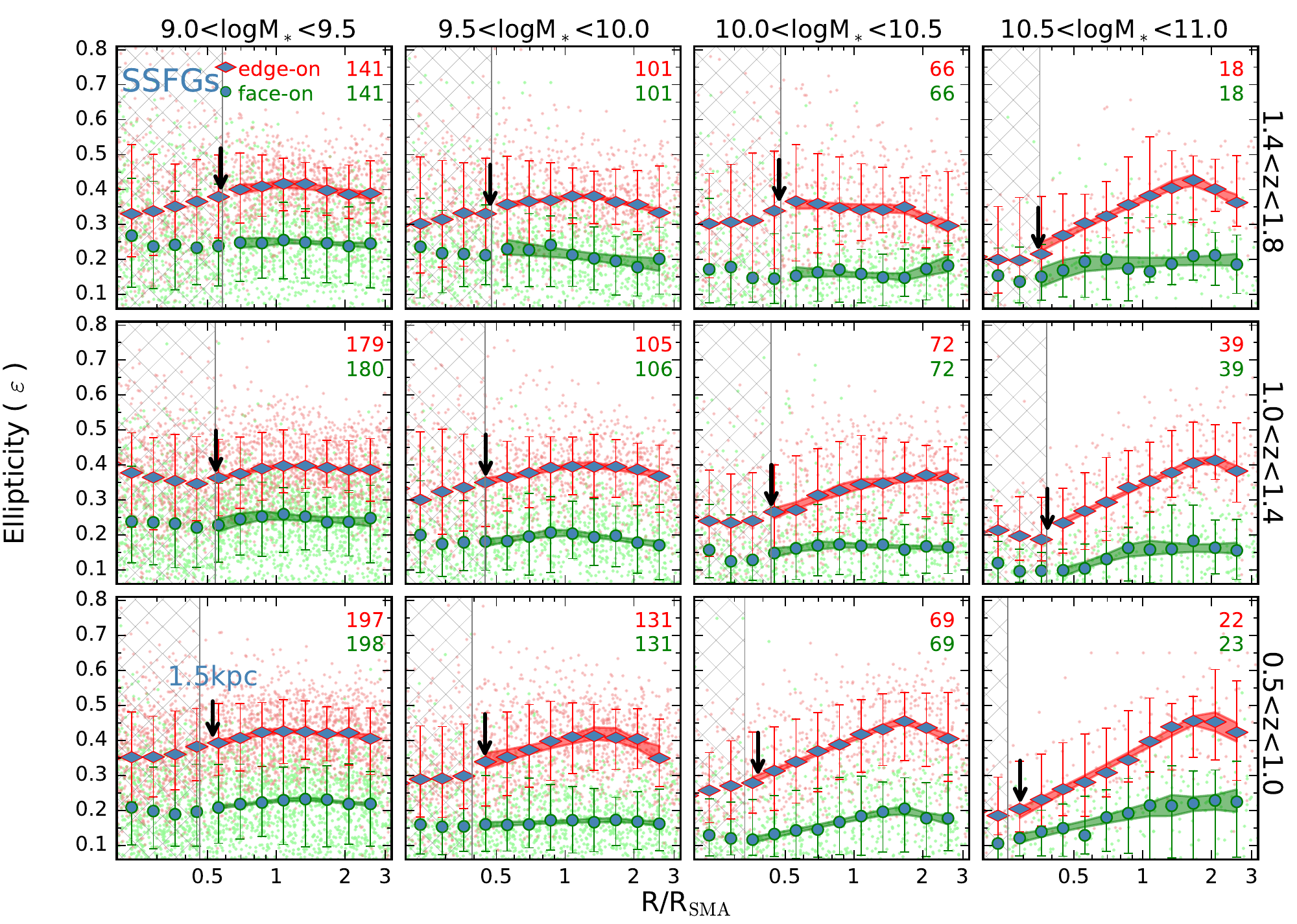}
\caption{
The composite ellipticity profiles as a function of normalised radius for
the SSFGs in classified mass-redshift bins. 
The steel blue diamonds with red edge represent the median values of 
ellipticities at every given radius for edge-on systems, followed by 
68\% confidence intervals. The steel blue circles with green edge 
represent the median values of ellipticities at every given radius 
for face-on systems, followed by 68\% confidence intervals. 
The shade regions indicate the ranges affected significantly by 
PSF smoothing (0.18\arcsec). The galaxy number of each class is shown 
on the right-top corner of each panel. 
The thick black arrows indicate the position of 
$\rm R_{SMA}=1.5kpc$ in each bin. The observed data are shown with tiny dots 
for edge-on (red) and face-on (green) systems, respectively. 
The shade regions show the best segmented models with $2\sigma$ lower and upper limits
for edge-on (red) and face-on (green) systems, respectively.  
\label{fig7}}
\end{figure*}

\begin{figure*}
\centering
\includegraphics[angle=0,width=\textwidth]{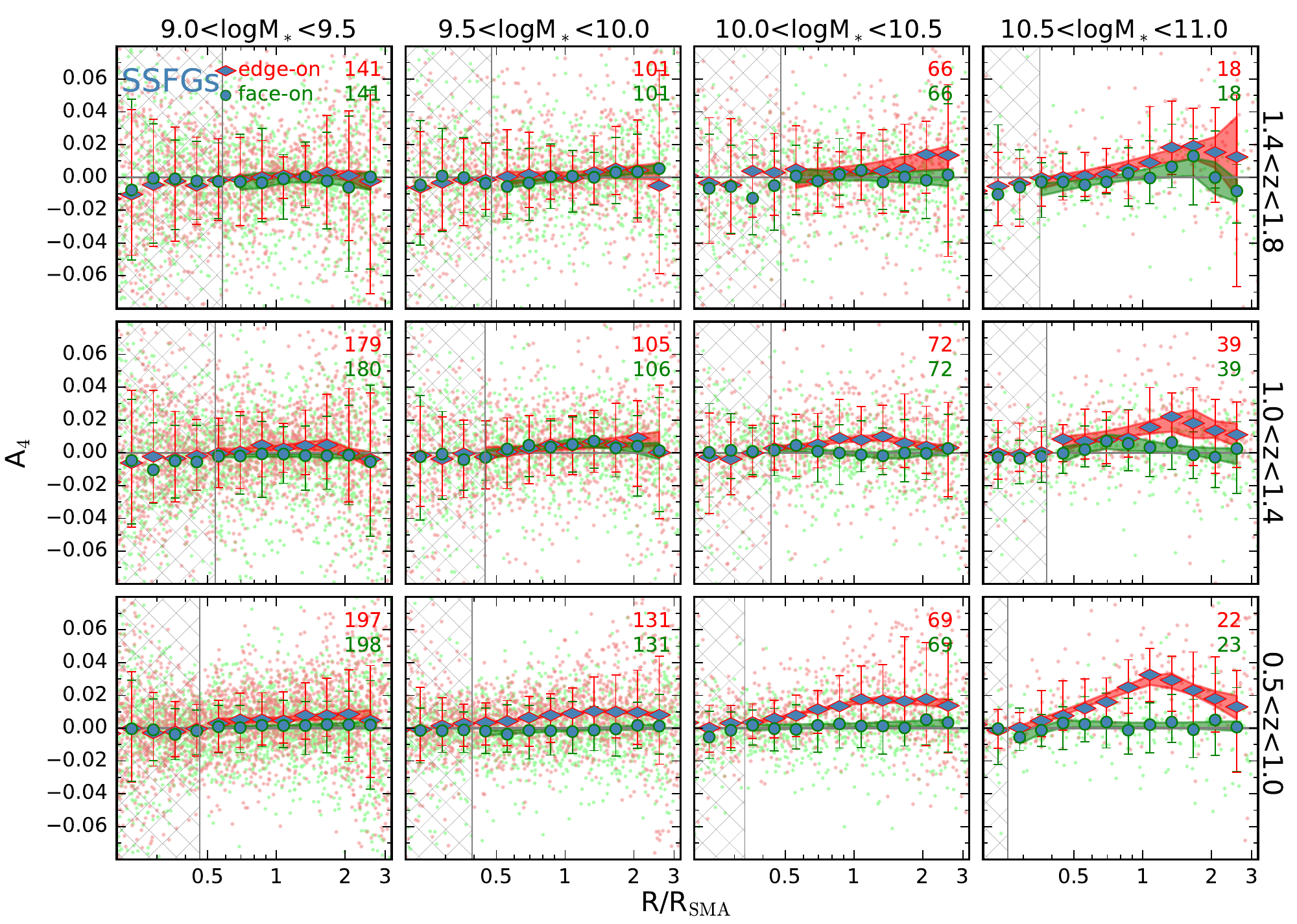}
\caption{
The composite A$_4$ as a function of normalised radius for the SSFGs. 
The data points, error-bars and lines have the same meanings as those in Figure 5. 
A$_4>0$ indicates disky, whereas A$_4<0$ indicates boxy. 
\label{fig8}}
\end{figure*}

\begin{figure*}
\centering
\includegraphics[angle=0,width=\textwidth]{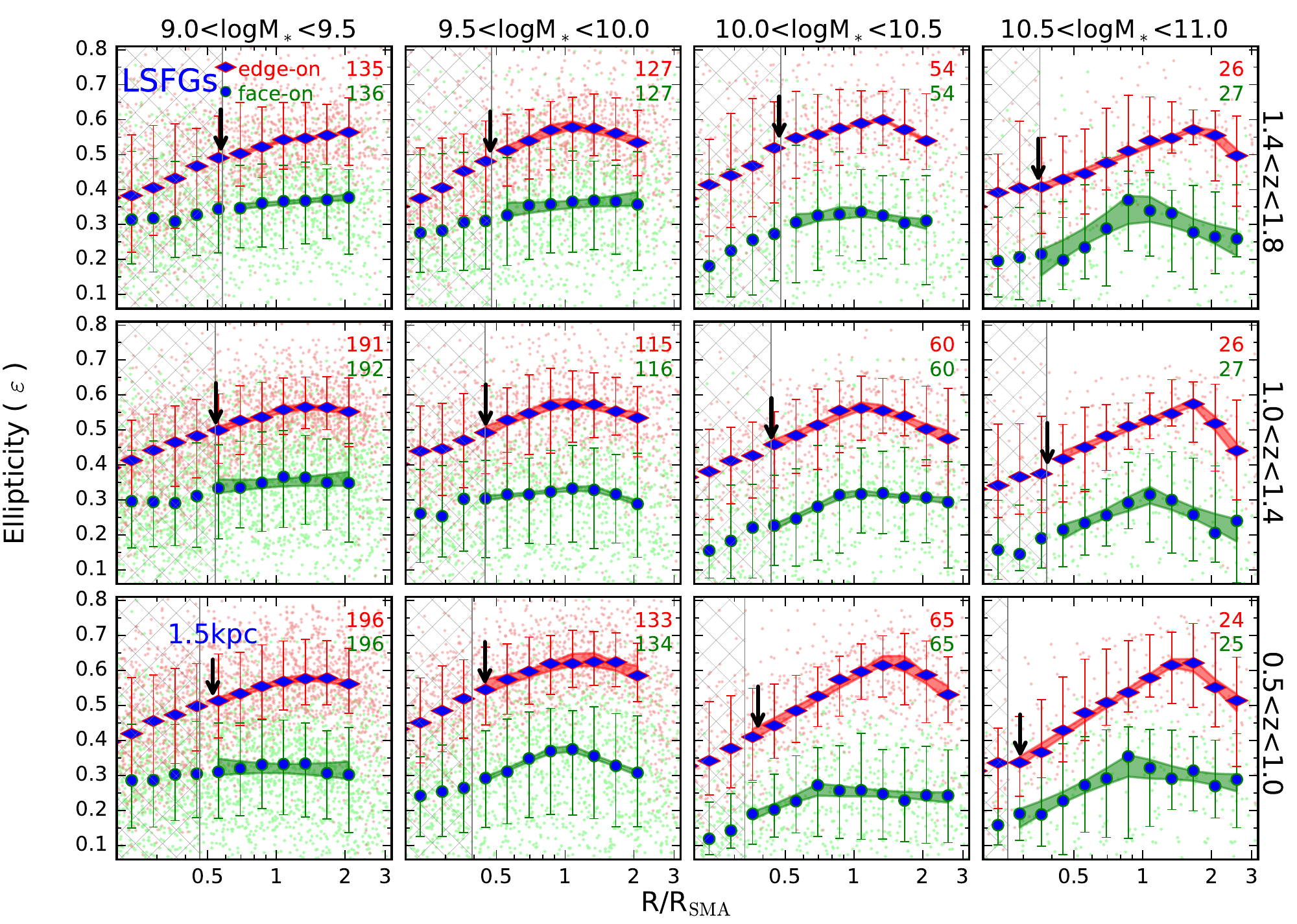}
\caption{
The composite ellipticity profiles as a function of normalised radius for
the LSFGs in classified mass-redshift bins.
The blue diamonds with red edge represent the median values of
ellipticities at every given radius for edge-on systems, followed by
68\% confidence intervals. The blue circles with green edge
represent the median values of ellipticities at every given radius
for face-on systems, followed by 68\% confidence intervals.
The shade regions indicate the ranges affected significantly by
PSF smoothing (0.18\arcsec). The galaxy number of each class is shown
on the right-top corner of each panel.
The thick black arrows indicate the position of
$\rm R_{SMA}=1.5kpc$ in each bin. The observed data are shown with tiny dots
for edge-on (red) and face-on (green) systems, respectively.
The shade regions show the best segmented models with $2\sigma$ lower and upper limits
for edge-on (red) and face-on (green) systems, respectively.
\label{fig5}}
\end{figure*}

\begin{figure*}
\centering
\includegraphics[angle=0,width=\textwidth]{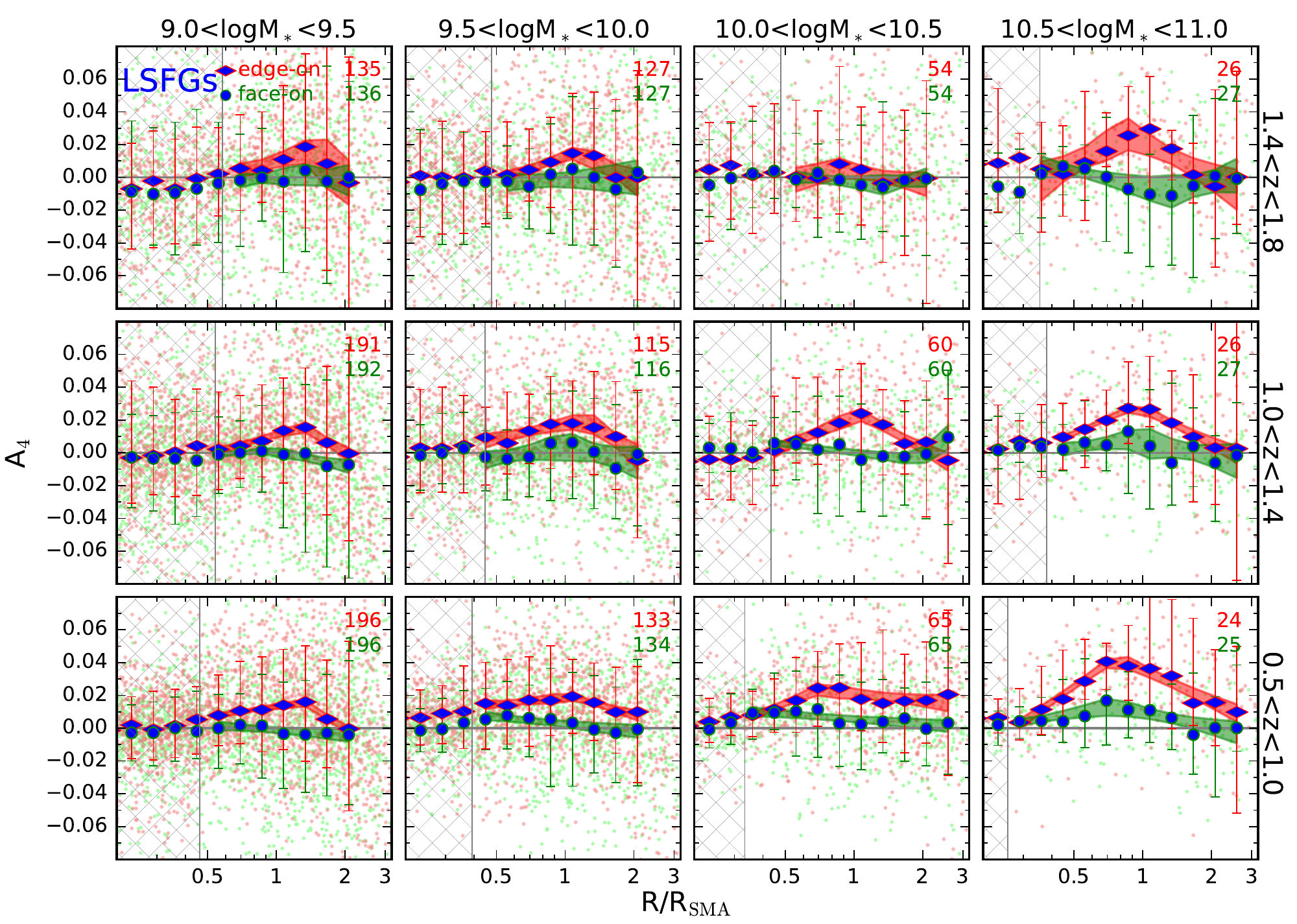}
\caption{
The composite A$_4$ as a function of normalised radius for the LSFGs. 
The data points, error-bars and lines have the same meanings as those in Figure 7. 
A$_4>0$ indicates disky, whereas A$_4<0$ indicates boxy.
\label{fig6}}
\end{figure*}

\begin{figure*}
\centering
\includegraphics[angle=0,width=\textwidth]{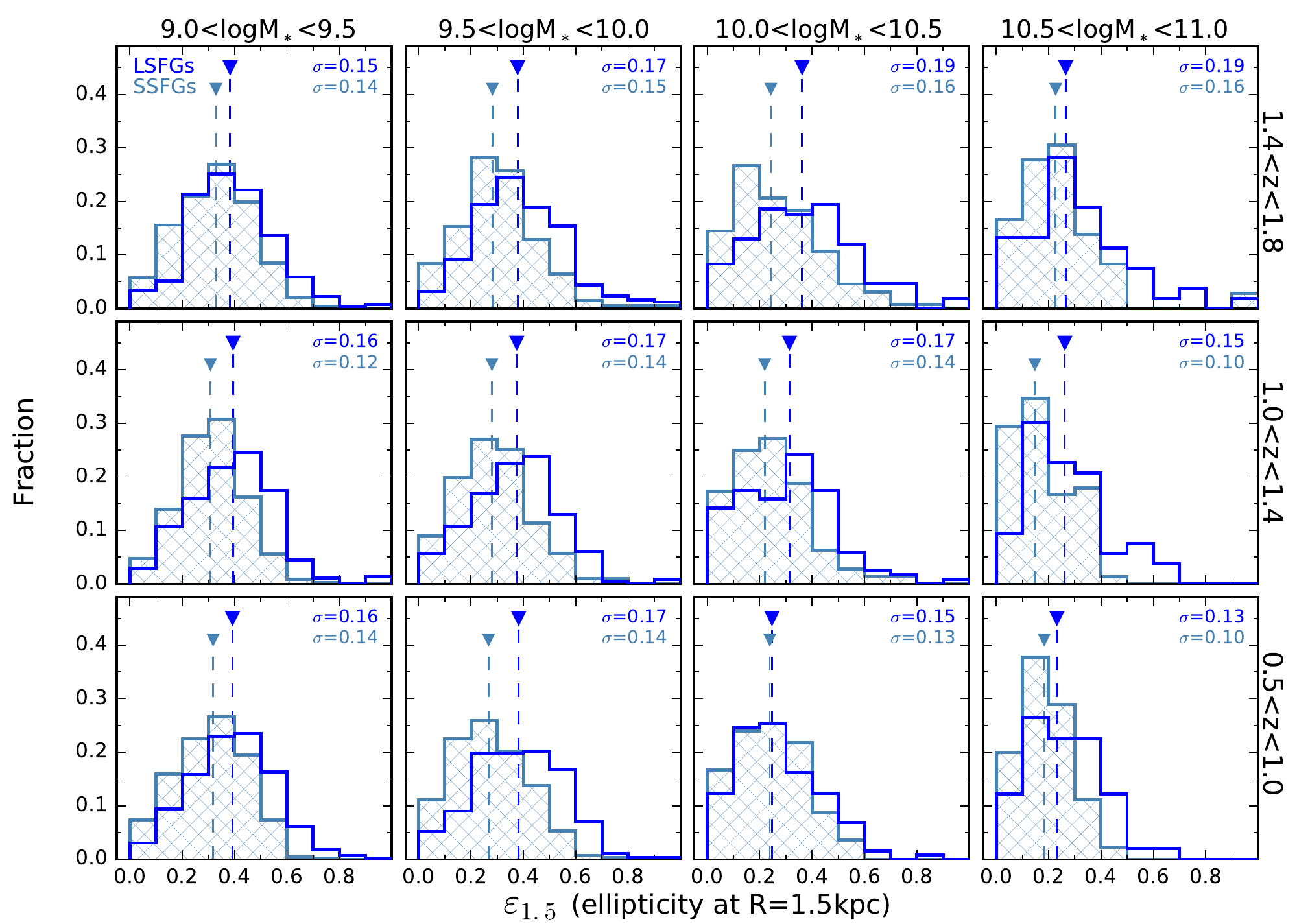}
\caption{
Distributions of ellipticity at the position of $\rm R_{SMA}=1.5kpc$
($\rm \varepsilon_{1.5}$) for LSFGs (blue) and SSFGs (steel blue) in each mass-redshift bin.
The median values are indicated with triangles plus dashed lines. The standard deviations
($\sigma$) are presented in the right-top corner of each panel.
\label{fig9}}
\end{figure*}

\begin{figure*}
\centering
\includegraphics[angle=0,width=\textwidth]{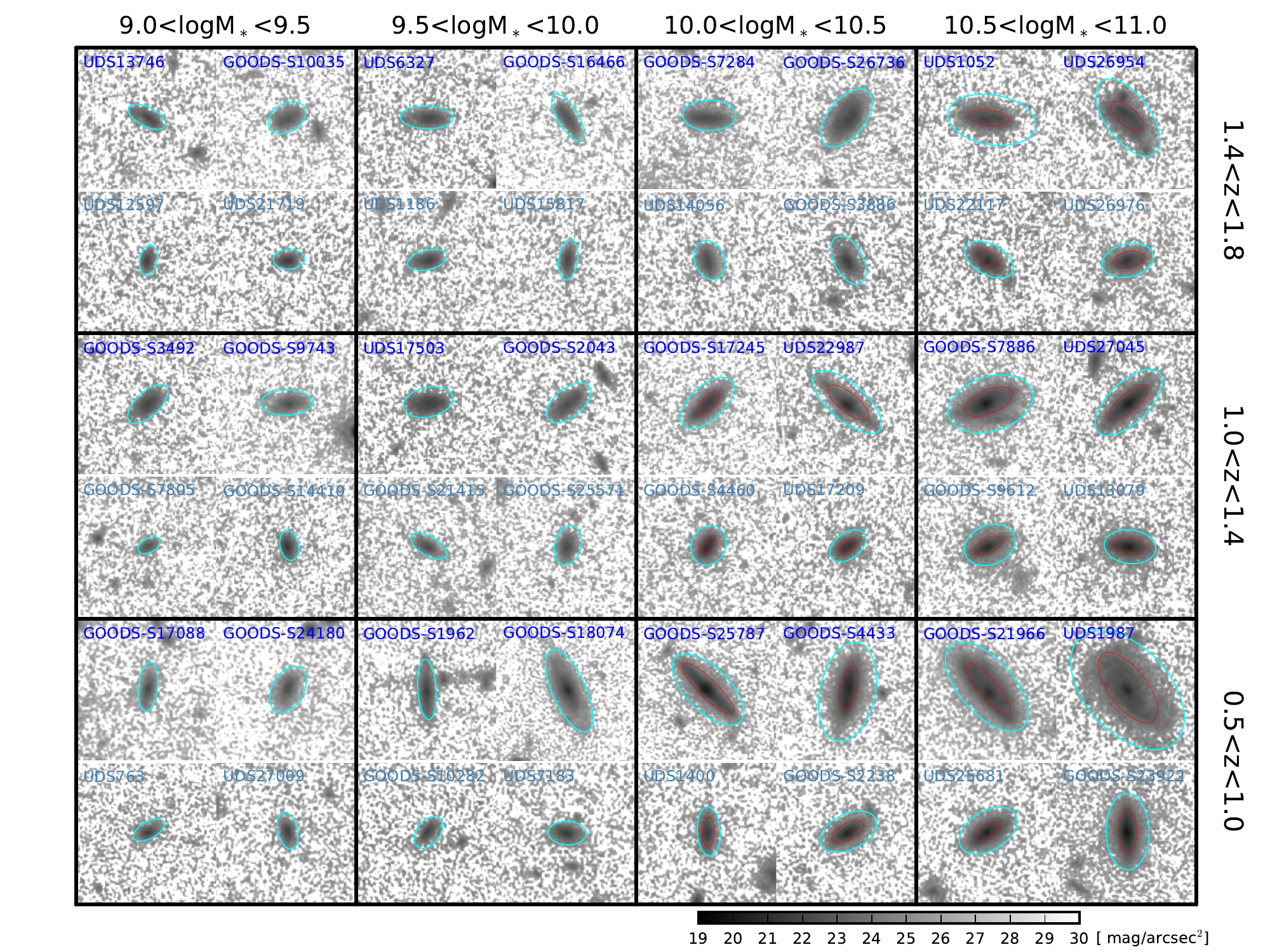}
\caption{
Cut-out images of example edge-on galaxies in F160W (for $1.0<z<1.8$) and
F125W (for $0.5<z<1.0$) in every mass-redshift bin.
The size of each cut-out image
is $50 \times 50$ kpc. The cyan ellipse in each cut-out image
indicates the outermost isophote we measured, which is close to 
the position of $\rm \sim 3R_{SMA}$. The red ellipses
indicate the positions of maximum ellipticity in the $\rm \varepsilon$ profiles of 
some galaxies with diffuse halo-like stellar components. 
In each bin, the upper images are for LSFGs, whereas the lower ones are for SSFGs.
\label{fig11}}
\end{figure*}

\begin{figure*}
\centering
\includegraphics[angle=0,width=\textwidth]{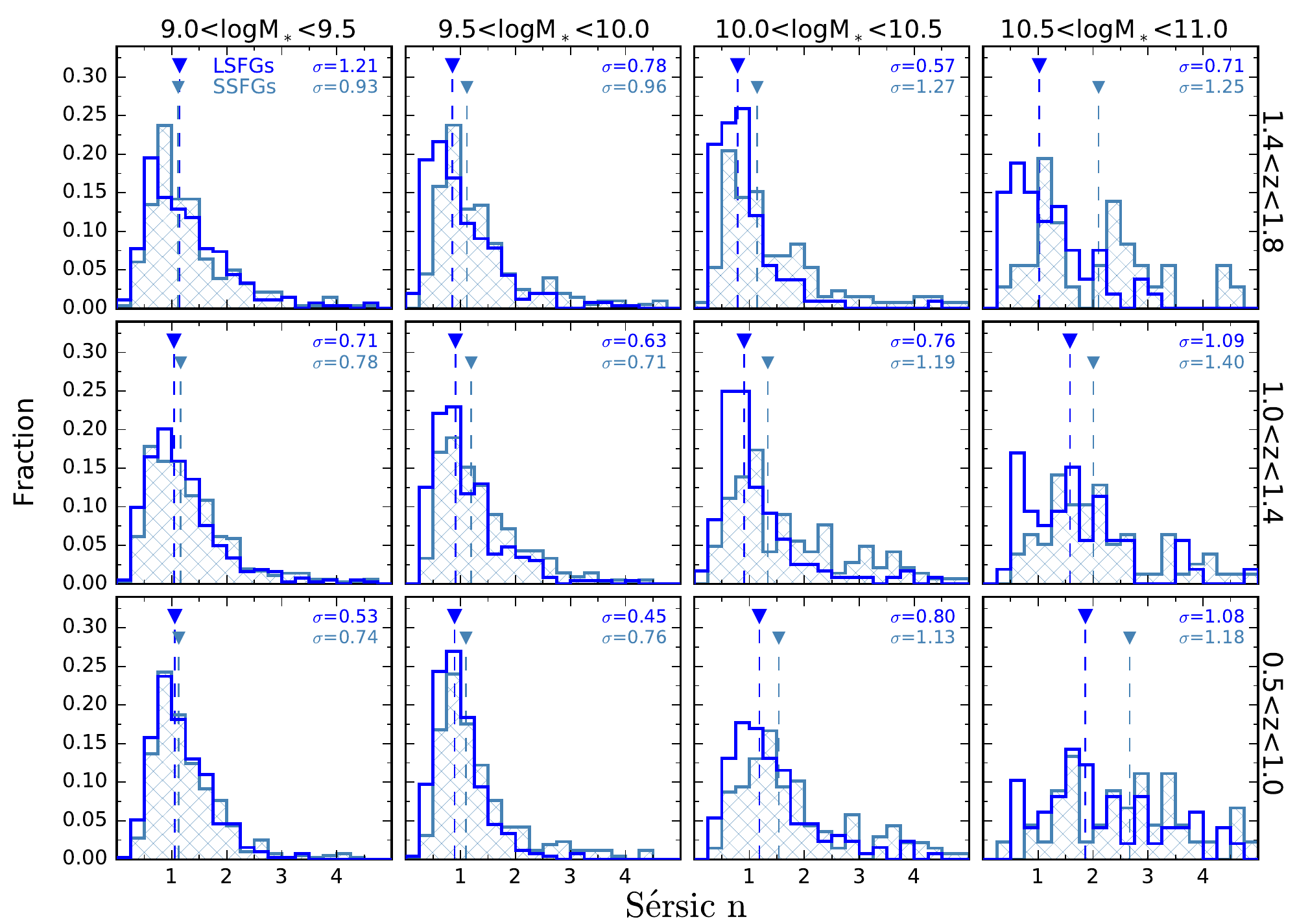}
\caption{
Distributions of S\'ersic Index $n$ for the LSFGs (blue) and 
SSFGs (steel blue) in every mass-redshift bin, respectively. 
The median values are indicated with triangles plus dashed lines. 
The standard deviations ($\sigma$) are presented in the right-top corner of each panel.
\label{fig10}}
\end{figure*}

\clearpage

\begin{table}[ht]
\begin{center}
\caption{Sample Selection Cuts}
\begin{tabular}{lcccccccc}
\hline\hline
{\footnotesize Cut} & {\footnotesize GOODS-S} &
{\footnotesize UDS} & {\footnotesize Combined}\\
\hline
Full catalog         & 34,930 (100.0\%) &35,932 (100.0\%) & 70,862 (100.0\%)\\
F160W($H$) $<$ 24.5  & 8,293 (23.7\%) & 9,671 (26.9\%) & 17,964 (25.4\%)\\
SE PhotFlag = 0      & 8,104 (23.2\%) & 9,151 (25.5\%) & 17,255 (24.4\%)\\
SE $\tt CLASS\_STAR <$ 0.9 & 7,891 (22.6\%) & 8,933 (24.9\%) & 16,824 (23.7\%)\\
$0.5 < z < 1.8$         & 4,753  (13.6\%) & 5,735  (16.0\%) & 10,488  (14.8\%)\\
$9.0 < log M* < 11.0$   & 3,261  (9.3\%)  & 3,964  (11.0\%) &  7,225  (10.2\%)\\
{\tt GALFIT} flag = 0   & 2,868  (8.2\%)  & 3,555  (9.9\%)  &  6,423  (9.1\%)\\
ISO PhotFlag = 0        & 2,690  (7.7\%)  & 3,434  (9.6\%)  &  6,124  (8.6\%)\\
$\rm R_{SMA}>0.18{\arcsec}$  & 2,250 (6.4\%)  & 2,829  (7.9\%) &5,079 (7.2\%)\\
$UVJ$-defined SFGs       & 2,036  (5.8\%)  & 2,571  (7.2\%) &  4,607  (6.5\%)\\
non-compact SFGs         & 2,033  (5.8\%)  & 2,562  (7.1\%) &  4,595  (6.5\%)\\
\hline
\end{tabular}
\end{center}
\end{table}

\begin{table}[ht]
\begin{center}
\caption{Parameters of the best linear fits to the size-mass relations}
\begin{tabular}{ccccccccc}
\hline\hline
{\footnotesize Redshift Range} & {\footnotesize Slope $a$} &
{\footnotesize Zeropoint $b$}\\
\hline
$0.5<z<1.0$ & $0.155$ & $-0.978$ \\
$1.0<z<1.4$ & $0.133$ & $-0.793$ \\
$1.4<z<1.8$ & $0.116$ & $-0.651$ \\
\hline
\end{tabular}
\end{center}
\end{table}

\begin{longrotatetable}
\begin{table}[ht]
\tablenum{3}
\tiny
\begin{center}
\caption{Best-fit parameters for the ellipticity and A$_4$ profiles}
\begin{tabular}{cccccccccccccc}
\hline\hline
\colhead{} & \colhead{} &
\multicolumn4c{$\rm 1.4<z<1.8$} &
\multicolumn4c{$\rm 1.0<z<1.4$} &
\multicolumn4c{$\rm 0.5<z<1.0$} \\
\cline{3-6}
\cline{11-14}
\colhead{Stellar Mass} & \colhead{Sub-class} &
\colhead{Slope} & \colhead{Intercept} & \colhead{R-squared} & \colhead{Break $\widetilde{R}$\tablenotemark{*} } &
\colhead{Slope} & \colhead{Intercept} & \colhead{R-squared} & \colhead{Break $\widetilde{R}$\tablenotemark{*} } &                                                 
\colhead{Slope} & \colhead{Intercept} & \colhead{R-squared} & \colhead{Break $\widetilde{R}$\tablenotemark{*} } \\
\hline                                                                                                           
\colhead{} & \colhead{} & \multicolumn8c{Best-fit parameters of ellipticity profiles for SSFGs (Figure 5)} \\
\hline
$\rm 9.0<logM_*<9.5$ & face-on & 0.033$\pm$0.098 & 0.252$\pm$0.011 & 0.59 & 1.06 & 0.184$\pm$0.133 & 0.275$\pm$0.027 & 0.60 & 0.80 & 0.070$\pm$0.014 & 0.227$\pm$0.002 & 0.92 & 1.26  \\
{} & {} & -0.033$\pm$0.016 & 0.254$\pm$0.017 & \nodata & \nodata & -0.039$\pm$0.023 & 0.253$\pm$0.031 & \nodata & \nodata & -0.057$\pm$0.014 & 0.240$\pm$0.002 & \nodata & \nodata  \\
{} & edge-on & 0.085$\pm$0.052 & 0.414$\pm$0.005 & 0.87 & 1.12 & 0.141$\pm$0.017 & 0.398$\pm$0.003 & 0.98 & 1.03 & 0.160$\pm$0.036 & 0.433$\pm$0.006 & 0.91 & 0.96  \\
{} & {} & -0.101$\pm$0.031 & 0.424$\pm$0.007 & \nodata & \nodata & -0.037$\pm$0.007 & 0.400$\pm$0.003 & \nodata & \nodata & -0.045$\pm$0.015 & 0.429$\pm$0.005 & \nodata & \nodata  \\
$\rm 9.5<logM_*<10.0$ & face-on & -0.075$\pm$0.021 & 0.216$\pm$0.005 & 0.69 & \nodata & 0.095$\pm$0.029 & 0.211$\pm$0.007 & 0.86 & 0.96 & 0.025$\pm$0.012 & 0.168$\pm$0.002 & 0.64 & 1.60  \\
{} & {} & \nodata & \nodata & \nodata & \nodata & -0.085$\pm$0.018 & 0.208$\pm$0.009 & \nodata & \nodata & -0.053$\pm$0.026 & 0.184$\pm$0.001 & \nodata & \nodata  \\
{} & edge-on & 0.080$\pm$0.018 & 0.377$\pm$0.002 & 0.96 & 1.25 & 0.121$\pm$0.019 & 0.395$\pm$0.003 & 0.92 & 1.18 & 0.164$\pm$0.035 & 0.399$\pm$0.006 & 0.87 & 1.51  \\
{} & {} & -0.149$\pm$0.017 & 0.400$\pm$0.002 & \nodata & \nodata & -0.091$\pm$0.025 & 0.411$\pm$0.003 & \nodata & \nodata & -0.287$\pm$0.089 & 0.480$\pm$0.004 & \nodata & \nodata  \\
$\rm 10.0<logM_*<10.5$ & face-on & -0.029$\pm$0.022 & 0.157$\pm$0.003 & 0.71 & 1.67 & 0.110$\pm$0.053 & 0.188$\pm$0.014 & 0.62 & 0.73 & 0.137$\pm$0.012 & 0.176$\pm$0.003 & 0.97 & 1.57  \\
{} & {} & 0.196$\pm$0.147 & 0.107$\pm$0.003 & \nodata & \nodata & -0.019$\pm$0.013 & 0.170$\pm$0.012 & \nodata & \nodata & -0.147$\pm$0.050 & 0.232$\pm$0.002 & \nodata & \nodata  \\
{} & edge-on & -0.032$\pm$0.028 & 0.351$\pm$0.004 & 0.92 & 1.66 & 0.234$\pm$0.043 & 0.344$\pm$0.010 & 0.97 & 1.01 & 0.265$\pm$0.012 & 0.405$\pm$0.003 & 0.99 & 1.61  \\
{} & {} & -0.252$\pm$0.056 & 0.399$\pm$0.004 & \nodata & \nodata & 0.063$\pm$0.029 & 0.344$\pm$0.009 & \nodata & \nodata & -0.255$\pm$0.045 & 0.513$\pm$0.001 & \nodata & \nodata  \\
$\rm 10.5<logM_*<11.0$ & face-on & 0.250$\pm$0.201 & 0.260$\pm$0.080 & 0.64 & 0.51 & 0.216$\pm$0.043 & 0.169$\pm$0.011 & 0.93 & 0.94 & 0.144$\pm$0.024 & 0.198$\pm$0.009 & 0.92 & 1.26  \\
{} & {} & 0.016$\pm$0.024 & 0.192$\pm$0.043 & \nodata & \nodata & -0.003$\pm$0.040 & 0.163$\pm$0.015 & \nodata & \nodata & 0.048$\pm$0.088 & 0.208$\pm$0.009 & \nodata & \nodata  \\
{} & edge-on & 0.305$\pm$0.020 & 0.373$\pm$0.005 & 0.99 & 1.58 & 0.300$\pm$0.012 & 0.343$\pm$0.002 & 0.99 & 1.89 & 0.346$\pm$0.015 & 0.379$\pm$0.004 & 0.99 & 1.83  \\
{} & {} & -0.308$\pm$0.074 & 0.495$\pm$0.003 & \nodata & \nodata & -0.318$\pm$0.097 & 0.514$\pm$0.002 & \nodata & \nodata & -0.316$\pm$0.205 & 0.553$\pm$0.006 & \nodata & \nodata  \\
\hline
\colhead{} & \colhead{} & \multicolumn8c{Best-fit parameters for the A$_4$ profiles of SSFGs (Figure 6)} \\
\hline
$\rm 9.0<logM_*<9.5$ & face-on & 0.015$\pm$0.035 & -0.001$\pm$0.004 & 0.53 & 1.05 & -0.001$\pm$0.002 & -0.002$\pm$0.000 & 0.53 & \nodata & 0.003$\pm$0.001 & 0.001$\pm$0.000 & 0.68 & 1.78  \\
{} & {} & -0.005$\pm$0.009 & -0.000$\pm$0.003 & \nodata & \nodata & \nodata & \nodata & \nodata & \nodata & -0.002$\pm$0.013 & 0.003$\pm$0.000 & \nodata & \nodata  \\
{} & edge-on & 0.008$\pm$0.004 & 0.000$\pm$0.000 & 0.65 & 1.86 & 0.011$\pm$0.005 & 0.003$\pm$0.001 & 0.78 & 1.59 & 0.007$\pm$0.002 & 0.006$\pm$0.000 & 0.55 & \nodata  \\
{} & {} & -0.034$\pm$0.034 & 0.012$\pm$0.000 & \nodata & \nodata & -0.045$\pm$0.017 & 0.014$\pm$0.001 & \nodata & \nodata & \nodata & \nodata & \nodata & \nodata  \\
$\rm 9.5<logM_*<10.0$ & face-on & 0.014$\pm$0.003 & -0.001$\pm$0.000 & 0.84 & \nodata & 0.018$\pm$0.006 & 0.005$\pm$0.001 & 0.74 & 1.19 & 0.005$\pm$0.001 & -0.001$\pm$0.000 & 0.67 & \nodata  \\
{} & {} & \nodata & \nodata & \nodata & \nodata & -0.017$\pm$0.011 & 0.008$\pm$0.002 & \nodata & \nodata & \nodata & \nodata & \nodata & \nodata  \\
{} & edge-on & 0.006$\pm$0.005 & 0.001$\pm$0.001 & 0.52 & \nodata & 0.011$\pm$0.004 & 0.004$\pm$0.001 & 0.56 & \nodata & 0.015$\pm$0.001 & 0.009$\pm$0.000 & 0.99 & 1.44  \\
{} & {} & \nodata & \nodata & \nodata & \nodata & \nodata & \nodata & \nodata & \nodata & -0.010$\pm$0.005 & 0.012$\pm$0.000 & \nodata & \nodata  \\
$\rm 10.0<logM_*<10.5$ & face-on & -0.002$\pm$0.005 & 0.001$\pm$0.001 & 0.52 & \nodata & -0.009$\pm$0.003 & -0.001$\pm$0.001 & 0.66 & 1.37 & 0.003$\pm$0.002 & 0.002$\pm$0.001 & 0.58 & \nodata  \\
{} & {} & \nodata & \nodata & \nodata & \nodata & 0.013$\pm$0.014 & -0.004$\pm$0.000 & \nodata & \nodata & \nodata & \nodata & \nodata & \nodata  \\
{} & edge-on & 0.020$\pm$0.007 & 0.004$\pm$0.001 & 0.59 & \nodata & 0.016$\pm$0.003 & 0.008$\pm$0.001 & 0.90 & 1.27 & 0.028$\pm$0.002 & 0.016$\pm$0.001 & 0.98 & 1.19  \\
{} & {} & \nodata & \nodata & \nodata & \nodata & -0.025$\pm$0.006 & 0.012$\pm$0.000 & \nodata & \nodata & -0.008$\pm$0.007 & 0.018$\pm$0.001 & \nodata & \nodata  \\
$\rm 10.5<logM_*<11.0$ & face-on & 0.021$\pm$0.006 & 0.002$\pm$0.001 & 0.69 & 1.71 & 0.041$\pm$0.019 & 0.013$\pm$0.005 & 0.63 & 0.70 & 0.049$\pm$0.024 & 0.021$\pm$0.011 & 0.72 & 0.43  \\
{} & {} & -0.082$\pm$0.064 & 0.025$\pm$0.001 & \nodata & \nodata & -0.016$\pm$0.009 & 0.005$\pm$0.007 & \nodata & \nodata & -0.002$\pm$0.003 & 0.002$\pm$0.007 & \nodata & \nodata  \\
{} & edge-on & 0.029$\pm$0.005 & 0.009$\pm$0.001 & 0.86 & 1.80 & 0.026$\pm$0.008 & 0.015$\pm$0.002 & 0.71 & 1.49 & 0.052$\pm$0.004 & 0.028$\pm$0.001 & 0.98 & 1.19  \\
{} & {} & -0.028$\pm$0.140 & 0.024$\pm$0.002 & \nodata & \nodata & -0.035$\pm$0.032 & 0.025$\pm$0.001 & \nodata & \nodata & -0.056$\pm$0.017 & 0.036$\pm$0.001 & \nodata & \nodata  \\
\hline
\end{tabular}
\end{center}
\end{table}
\end{longrotatetable}

\begin{longrotatetable}
\begin{table}[ht]
\tablenum{3}
\tiny
\begin{center}
\caption{(continued)}
\begin{tabular}{cccccccccccccc}
\hline
\hline
\colhead{} & \colhead{} &
\multicolumn4c{$\rm 1.4<z<1.8$} &
\multicolumn4c{$\rm 1.0<z<1.4$} &
\multicolumn4c{$\rm 0.5<z<1.0$} \\
\cline{3-6}
\cline{11-14}
\colhead{Stellar Mass} & \colhead{Sub-class} &
\colhead{Slope} & \colhead{Intercept} & \colhead{R-squared} & \colhead{Break $\widetilde{R}$\tablenotemark{*} } &
\colhead{Slope} & \colhead{Intercept} & \colhead{R-squared} & \colhead{Break $\widetilde{R}$\tablenotemark{*} } &                                                 
\colhead{Slope} & \colhead{Intercept} & \colhead{R-squared} & \colhead{Break $\widetilde{R}$\tablenotemark{*} } \\
\hline
\colhead{} & \colhead{} & \multicolumn8c{Best-fit parameters of ellipticity profiles for LSFGs (Figure 7)} \\
\hline
$\rm 9.0<logM_*<9.5$ & face-on & 0.055$\pm$0.010 & 0.361$\pm$0.002 & 0.87 & \nodata & 0.037$\pm$0.023 & 0.349$\pm$0.004 & 0.54 & \nodata & -0.017$\pm$0.028 & 0.320$\pm$0.005 & 0.07 & \nodata  \\
{} & {} & \nodata & \nodata & \nodata & \nodata & \nodata & \nodata & \nodata & \nodata & \nodata & \nodata & \nodata & \nodata  \\
{} & edge-on & 0.201$\pm$0.033 & 0.535$\pm$0.004 & 1.00 & 1.07 & 0.195$\pm$0.023 & 0.552$\pm$0.003 & 0.98 & 1.23 & 0.195$\pm$0.025 & 0.564$\pm$0.004 & 0.98 & 1.23  \\
{} & {} & 0.075$\pm$0.008 & 0.539$\pm$0.003 & \nodata & \nodata & -0.065$\pm$0.035 & 0.575$\pm$0.004 & \nodata & \nodata & -0.078$\pm$0.034 & 0.589$\pm$0.002 & \nodata & \nodata  \\
$\rm 9.5<logM_*<10.0$ & face-on & 0.053$\pm$0.023 & 0.356$\pm$0.004 & 0.51 & \nodata & 0.069$\pm$0.014 & 0.330$\pm$0.003 & 0.95 & 1.27 & 0.283$\pm$0.021 & 0.388$\pm$0.005 & 0.99 & 0.98  \\
{} & {} & \nodata & \nodata & \nodata & \nodata & -0.211$\pm$0.032 & 0.358$\pm$0.004 & \nodata & \nodata & -0.241$\pm$0.023 & 0.383$\pm$0.006 & \nodata & \nodata  \\
{} & edge-on & 0.307$\pm$0.059 & 0.589$\pm$0.011 & 0.96 & 1.00 & 0.263$\pm$0.040 & 0.589$\pm$0.009 & 0.95 & 0.96 & 0.200$\pm$0.043 & 0.624$\pm$0.008 & 0.89 & 1.21  \\
{} & {} & -0.153$\pm$0.033 & 0.589$\pm$0.013 & \nodata & \nodata & -0.136$\pm$0.033 & 0.582$\pm$0.011 & \nodata & \nodata & -0.201$\pm$0.082 & 0.656$\pm$0.010 & \nodata & \nodata  \\
$\rm 10.0<logM_*<10.5$ & face-on & 0.122$\pm$0.080 & 0.341$\pm$0.013 & 0.79 & 0.97 & 0.310$\pm$0.026 & 0.331$\pm$0.006 & 0.99 & 0.95 & 0.261$\pm$0.042 & 0.300$\pm$0.015 & 0.92 & 0.69  \\
{} & {} & -0.116$\pm$0.039 & 0.338$\pm$0.010 & \nodata & \nodata & -0.057$\pm$0.017 & 0.322$\pm$0.006 & \nodata & \nodata & -0.039$\pm$0.026 & 0.253$\pm$0.020 & \nodata & \nodata  \\
{} & edge-on & 0.152$\pm$0.010 & 0.584$\pm$0.001 & 1.00 & 1.31 & 0.284$\pm$0.026 & 0.559$\pm$0.005 & 0.97 & 1.15 & 0.372$\pm$0.023 & 0.581$\pm$0.005 & 0.98 & 1.47  \\
{} & {} & -0.310$\pm$0.015 & 0.639$\pm$0.002 & \nodata & \nodata & -0.279$\pm$0.044 & 0.594$\pm$0.006 & \nodata & \nodata & -0.446$\pm$0.088 & 0.718$\pm$0.004 & \nodata & \nodata  \\
$\rm 10.5<logM_*<11.0$ & face-on & 0.396$\pm$0.082 & 0.367$\pm$0.025 & 0.87 & 0.95 & 0.277$\pm$0.061 & 0.306$\pm$0.014 & 0.90 & 1.07 & 0.316$\pm$0.052 & 0.347$\pm$0.019 & 0.90 & 0.86  \\
{} & {} & -0.255$\pm$0.079 & 0.352$\pm$0.031 & \nodata & \nodata & -0.266$\pm$0.061 & 0.322$\pm$0.017 & \nodata & \nodata & -0.103$\pm$0.058 & 0.321$\pm$0.027 & \nodata & \nodata  \\
{} & edge-on & 0.250$\pm$0.010 & 0.517$\pm$0.002 & 0.99 & 1.85 & 0.262$\pm$0.016 & 0.520$\pm$0.003 & 0.98 & 1.73 & 0.422$\pm$0.016 & 0.567$\pm$0.005 & 0.99 & 1.50  \\
{} & {} & -0.598$\pm$0.127 & 0.744$\pm$0.001 & \nodata & \nodata & -0.812$\pm$0.154 & 0.776$\pm$0.003 & \nodata & \nodata & -0.562$\pm$0.088 & 0.739$\pm$0.005 & \nodata & \nodata  \\
\hline
\colhead{} & \colhead{} & \multicolumn8c{Best-fit parameters for the A$_4$ profiles of LSFGs (Figure 8)} \\
\hline
$\rm 9.0<logM_*<9.5$ & face-on & 0.012$\pm$0.021 & 0.000$\pm$0.002 & 0.59 & 1.32 & 0.011$\pm$0.007 & 0.002$\pm$0.001 & 0.86 & 0.93 & -0.010$\pm$0.004 & -0.001$\pm$0.001 & 0.61 & \nodata  \\
{} & {} & -0.018$\pm$0.033 & 0.004$\pm$0.001 & \nodata & \nodata & -0.028$\pm$0.010 & 0.001$\pm$0.002 & \nodata & \nodata & \nodata & \nodata & \nodata & \nodata  \\
{} & edge-on & 0.042$\pm$0.018 & 0.010$\pm$0.002 & 0.80 & 1.42 & 0.040$\pm$0.006 & 0.011$\pm$0.001 & 0.96 & 1.31 & 0.021$\pm$0.004 & 0.013$\pm$0.001 & 0.97 & 1.32  \\
{} & {} & -0.119$\pm$0.108 & 0.035$\pm$0.001 & \nodata & \nodata & -0.086$\pm$0.016 & 0.026$\pm$0.001 & \nodata & \nodata & -0.089$\pm$0.011 & 0.027$\pm$0.000 & \nodata & \nodata  \\
$\rm 9.5<logM_*<10.0$ & face-on & 0.005$\pm$0.009 & -0.002$\pm$0.002 & 0.55 & \nodata & 0.023$\pm$0.018 & 0.004$\pm$0.005 & 0.54 & 1.01 & -0.014$\pm$0.003 & 0.002$\pm$0.001 & 0.75 & \nodata  \\
{} & {} & \nodata & \nodata & \nodata & \nodata & -0.033$\pm$0.026 & 0.004$\pm$0.004 & \nodata & \nodata & \nodata & \nodata & \nodata & \nodata  \\
{} & edge-on & 0.049$\pm$0.010 & 0.013$\pm$0.002 & 0.90 & 1.13 & 0.029$\pm$0.010 & 0.018$\pm$0.002 & 0.86 & 1.24 & 0.013$\pm$0.008 & 0.019$\pm$0.002 & 0.87 & 1.07  \\
{} & {} & -0.071$\pm$0.029 & 0.019$\pm$0.001 & \nodata & \nodata & -0.102$\pm$0.029 & 0.030$\pm$0.003 & \nodata & \nodata & -0.039$\pm$0.008 & 0.020$\pm$0.003 & \nodata & \nodata  \\
$\rm 10.0<logM_*<10.5$ & face-on & -0.013$\pm$0.011 & -0.003$\pm$0.002 & 0.59 & 1.33 & -0.017$\pm$0.003 & 0.000$\pm$0.001 & 0.85 & 1.91 & -0.010$\pm$0.003 & 0.006$\pm$0.001 & 0.64 & \nodata  \\
{} & {} & 0.023$\pm$0.026 & -0.008$\pm$0.002 & \nodata & \nodata & 0.107$\pm$0.049 & -0.035$\pm$0.001 & \nodata & \nodata & \nodata & \nodata & \nodata & \nodata  \\
{} & edge-on & 0.036$\pm$0.052 & 0.008$\pm$0.011 & 0.51 & 0.86 & 0.060$\pm$0.012 & 0.022$\pm$0.003 & 0.94 & 1.05 & 0.056$\pm$0.008 & 0.032$\pm$0.003 & 0.92 & 0.69  \\
{} & {} & -0.025$\pm$0.016 & 0.004$\pm$0.015 & \nodata & \nodata & -0.075$\pm$0.011 & 0.025$\pm$0.004 & \nodata & \nodata & -0.014$\pm$0.008 & 0.021$\pm$0.003 & \nodata & \nodata  \\
$\rm 10.5<logM_*<11.0$ & face-on & -0.032$\pm$0.011 & -0.006$\pm$0.003 & 0.70 & 1.27 & 0.020$\pm$0.015 & 0.009$\pm$0.004 & 0.54 & 0.87 & 0.025$\pm$0.008 & 0.017$\pm$0.004 & 0.72 & 0.74  \\
{} & {} & 0.034$\pm$0.029 & -0.013$\pm$0.004 & \nodata & \nodata & -0.027$\pm$0.023 & 0.006$\pm$0.004 & \nodata & \nodata & -0.030$\pm$0.010 & 0.010$\pm$0.004 & \nodata & \nodata  \\
{} & edge-on & 0.076$\pm$0.030 & 0.031$\pm$0.009 & 0.75 & 0.86 & 0.062$\pm$0.006 & 0.030$\pm$0.001 & 0.97 & 0.94 & 0.097$\pm$0.006 & 0.054$\pm$0.002 & 0.98 & 0.76  \\
{} & {} & -0.071$\pm$0.021 & 0.022$\pm$0.005 & \nodata & \nodata & -0.068$\pm$0.009 & 0.027$\pm$0.001 & \nodata & \nodata & -0.064$\pm$0.008 & 0.036$\pm$0.003 & \nodata & \nodata  \\
\hline
\end{tabular}
\end{center}
\tablenotetext{*}{\tiny $\widetilde{R}=R/R_{SMA}$}
\end{table}
\end{longrotatetable}

\end{document}